\documentclass[10pt,twocolumn,amsmath,amssymb,aps,prx,superscriptaddress]{revtex4-2}
\usepackage{graphicx}
\usepackage{dcolumn}
\usepackage{bm}
\usepackage{epstopdf}
\usepackage{amsmath}
\usepackage{amssymb}
\usepackage{xcolor}
\usepackage{float}
\usepackage{natbib}
\usepackage[colorlinks=true,linkcolor=black]{hyperref} 

\newcommand{\SI}{Supplementary Information}
\newcommand{\SIM}{Supplementary Method}
\newcommand{\SIR}{Supplementary Results}

\begin{document}

\title{Iterative structural coarse-graining for contagion dynamics in complex networks}

\author{Leyang Xue}
\affiliation{International Academic Center of Complex Systems, Beijing Normal University, Zhuhai, 519087, China}
\affiliation{School of Systems Science, Beijing Normal University, Beijing, 100875, China}

\author{Zengru Di}
\affiliation{International Academic Center of Complex Systems, Beijing Normal University, Zhuhai, 519087, China}
\affiliation{School of Systems Science, Beijing Normal University, Beijing, 100875, China}

\author{An Zeng*\thanks{anzeng@bnu.edu.cn}}
\affiliation{School of Systems Science, Beijing Normal University, Beijing, 100875, China}
\date{\today}

\begin{abstract}
Contagion dynamics in complex networks drive critical phenomena such as epidemic spread and information diffusion,
but their analysis remains computationally prohibitive in large-scale, high-complexity systems.
Here, we introduce the Iterative Structural Coarse-Graining (ISCG) framework, a scalable methodology that reduces network complexity while preserving key contagion dynamics with high fidelity.
Importantly, we derive theoretical conditions ensuring the precise preservation of both macroscopic outbreak sizes and microscopic node-level infection probabilities during network reduction.
Under these conditions, extensive experiments on diverse empirical networks demonstrate that ISCG achieves significant complexity reduction without sacrificing   prediction accuracy.
Beyond simplification, ISCG reveals multiscale structural patterns that govern contagion processes, enabling practical solutions to longstanding challenges in contagion dynamics.
Specifically, ISCG outperforms traditional adaptive centrality-based approaches in identifying influential spreaders, immunizing critical edges, and optimizing sentinel placement for early outbreak detection, offering superior accuracy and computational efficiency.
By bridging computational efficiency with dynamical fidelity, ISCG provides a transformative framework for analyzing large-scale contagion processes, with broad applications for epidemiology, information dissemination, and network resilience.
\end{abstract}
\maketitle

\section{Introduction}
\label{sec:introduction}
Complex networks play a pivotal role in shaping the dynamics of contagion processes, underpinning the spread of infectious diseases~\cite{pastor2015,zhang2016,battiston2020,Pastor2001,colizza2006}, driving information diffusion~\cite{Iribarren2009,davis2020}, and influencing opinion formation~\cite{Nardini2008,schawe2022}.
Modeling and predicting these processes is critical for understanding their impacts across health, social, and technological domains.
However, the intrinsic complexity of large-scale networks poses significant computational challenges, as the cost of simulating contagion dynamics often scales polynomially with network size.
These challenges limit the feasibility of real-time analysis and decision-making, particularly in urgent scenarios such as epidemic response.
To address these challenges, numerous theoretical frameworks have been developed, including mean-field approximations~\cite{Pastor2001,Moreno2002,sharkey2008,Mieghem2009,Castellano2010}, generating functions~\cite{Newman2002,volz2008}, and message-passing methods~\cite{karrer2010,lokhov2015,cantwell2019}. 
While these approaches provide valuable insights, they often rely on simplifying assumptions that obscure critical structural details, limiting their accuracy on heterogeneous and realistic networks.

Network coarse-graining offers a promising approach to simplify complex networks into smaller, representative structures while preserving essential properties.
By balancing computational efficiency with fidelity to dynamic behaviors, it enables scalable analyses of processes such as synchronization. 
Over the years, a variety of coarse-graining techniques have been developed~\cite{sales2007,gfeller,Gfeller2007,Gfeller2008,song2005,zheng2020,thibeault2020}, including clustering-based methods~(e.g., modularity optimization~\cite{girvan2002}, spectral clustering~\cite{sales2007}, and K-means approaches~\cite{Shuang2016,zeng2019}), and renormalization techniques~(e.g., geographical coarse-graining~\cite{kim2004}, box-counting~\cite{song2005}, and geometric embedding~\cite{garcia2018,zheng2020}).
Recent advances, such as spectral coarse-graining~\cite{Gfeller2007,Gfeller2008} and dimension reduction approaches~\cite{laurence2019,thibeault2020,vegue2023dimension,ghosh2023dimension, gao2024intrinsic, thibeault2024low}, have focused on preserving dynamic properties for specific processes like random walks~\cite{Gfeller2007} and synchronization~\cite{Moon2006,Gfeller2008,zeng2011,chen2013}.

Despite notable advancements, existing coarse-graining methods face critical limitations.
Structural-based approaches, such as community detection, are computational efficiency but often fail to retain dynamic properties essential for understanding network behavior. 
Conversely, methods emphasizing dynamical fidelity, like spectral coarse-graining~\cite{Gfeller2007,Gfeller2008}, face three challenges: (1) their dependence on eigenvector computations makes them computationally expensive and impractical for very large networks;
(2) they are typically designed to preserve specific eigenvalue-driven properties, limiting their applicability to diverse processes, including nonlinear or higher-order contagion dynamics.
(3) While effective at capturing global structural features, they struggle to preserve localized dynamic processes that arise from the structural heterogeneity inherent to real-world networks.
This trade-off between computational efficiency and dynamic fidelity highlights a fundamental challenge in network science: the need for a scalable framework capable of reducing computational complexity while preserving the intricate interplay between structure and dynamics. 
Such a framework must also accommodate localized dynamics and structural diversity, ensuring meaningful and actionable insights across diverse network scales and configurations.

To address these challenges, we propose an iterative structural coarse-graining (ISCG) framework that leverages local structural information to achieve scalability while preserving dynamical fidelity.
Unlike spectral methods, ISCG avoids computationally expensive global eigenvector calculations, relying instead on an iterative process that systematically simplifies networks to any desired scale.
By aggregating $k$-clique structures into super-nodes, ISCG merges nodes with similar contagion characteristics, effectively capturing localized contagion processes.
This scalable approach is particularly well-suited for large and heterogeneous networks, preserving the intricate interplay between structure and dynamics while offering intuitive insights into multiscale contagion behavior.
By integrating simplification with theoretical rigor, ISCG reveals hidden structural patterns that drive contagion dynamics, enabling the development of actionable strategies.
These insights lead to practical solutions to three classical propagation challenges: maximizing influence, immunizing critical edges, and optimizing sentinel placement for early outbreak detection.
Comparative analyses demonstrate that ISCG-based strategies consistently outperform traditional adaptive centrality methods, highlighting their practicality and broad applicability in solving diverse contagion-related problems.

Beyond its methodological advancements, the ISCG framework establishes a robust theoretical foundation for balancing network reduction with the preservation of dynamic fidelity.
Through rigorous analysis, we derive critical transmission probability thresholds that ensure contagion dynamics are accurately retained during substantial network simplifications.
Using the susceptible-infected-recovered (SIR) model,  we validate that ISCG preserves both macroscopic outbreak sizes and microscopic node-level infection probabilities at these thresholds, even after significant reductions in network complexity.
Moreover, ISCG supports tunable trade-offs, enabling approximate reductions that achieve greater simplifications while maintaining high levels of accuracy.
This flexibility addresses the computational challenges inherent in large-scale systems, broadening the framework's applicability across diverse domains such as epidemic modeling, information diffusion, and infrastructure resilience.
By integrating theoretical rigor with practical flexibility, ISCG advances scalable network analysis and dynamic modeling.
Its capability to simplify complex networks while preserving essential dynamics provides a powerful toolset for addressing contagion-based challenges in real-world scenarios.

\section{Model and Framework}
\label{subsec:model}
\subsection*{Contagion Model}
We adopt the SIR model, a paradigmatic framework widely  used to study contagion dynamics in complex networks. 
Theoretically, the model has been extensively applied to describe the spread of phenomena such as epidemics, information, knowledge, and innovation~\cite{pastor2015, zhang2016}. 
Numerically, it serves as a versatile tool for simulating contagion processes in practical contexts, ranging from controlling epidemic outbreaks~\cite{chen2008, schneider2011} to optimizing the dissemination of brands or products~\cite{kitsak2010, xue2022}.

In the model, individuals are classified into three states: susceptible~($S$), infected~($I$), and recovered~($R$),
Susceptible individuals ($S$) can contract the contagion, infected individuals ($I$) actively spread it, and recovered individuals ($R$) have acquired permanent immunity.
In its discrete-time implementation, all nodes initially begin in the $S$ state, except for a small subset of seed nodes initialized in the $I$ state.
At each time step, infected nodes ($I$) transmit the contagion to their susceptible neighbors ($S$) with a probability~$\beta$~(the transmission probability).
Subsequently, infected nodes transition to the recovered ($R$) state with a recovery rate $\mu$.
The process continues until no infected nodes remain in the system.

A critical metric for characterizing the contagion process is the final infection density, defined as $\rho=\frac{1}{N}\sum_{i=1}^{N}s(v_i)$, where $s(v_i)=1$ if node $v_i$ becomes infected during the process and $s(v_i)=0$ otherwise. 
The parameter $\rho$ serves as the macroscopic order parameter, capturing the steady-state behavior of the SIR model and linking it to the geometric properties of bond percolation~\cite{Newman2002}.
For simplicity, and without loss of generality, we set the recovery rate $\mu = 1$ throughout this study. 
This assumption streamlines the analysis while maintaining the model’s general applicability.

\begin{figure*}[!htbp]
	\centering
	\includegraphics[width=\textwidth]{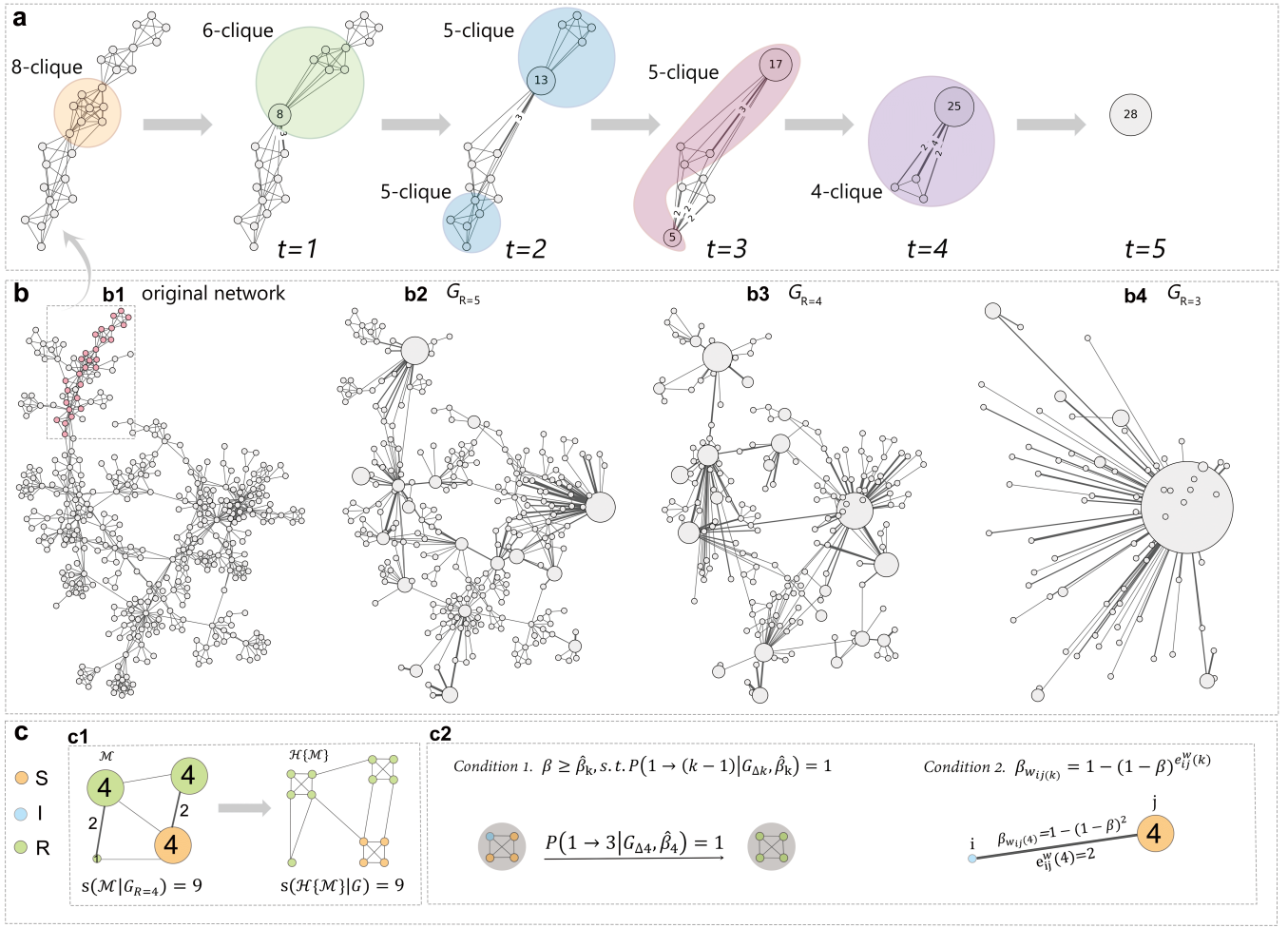}
		\caption{             
        \textbf{Illustration of the iterative structural coarse-graining~(ISCG) framework for preserving contagion dynamics.} 
        \textbf{a} Schematic of the coarse-graining method. 
        Maximal cliques in the network (e.g., orange circle) are merged into super-nodes.
        Each super-node inherits all links from the merged clique, and its weight is updated as the sum of the constituent nodes. 
        For multiple links between a super-node and another node, a single link is retained with a weight updated equal to the sum of the original links.
        Initially, all nodes and edges are assigned a weight of 1.
        This process produces a weighted network at each iteration step~($t=1$, second column), and the procedure continues until the network is reduced to a single node.
        \textbf{b} Construction of $k$-clique coarse-graining networks~(CGNs). 
        The ISCG method is applied iteratively to reduce the network until all cliques in the network are smaller than size $k$, resulting in a $k$-clique CGN~($G_{R=k}$). 
        The size of each node represents its weight, while the thickness of each link corresponds to the weight of the edge.
        \textbf{b1} Original network~(e.g., a co-authorship network).  
        \textbf{b2} 5-clique CGN. 
        \textbf{b3} 4-clique CGN.
        \textbf{b4} 3-clique CGN. 
        \textbf{c} Perserving the final outbreak size~($s$) in SIR dynamics.  
        The final outbreak size ($s$) in the original network is inferred from the contagion dynamics on the $k$-clique CGN.
        \textbf{c1} For a given contagion configuration~($\mathcal{M}$) on $G_{R=4}$, all nodes represented by a super-node are assumed to be infected if the super-node itself is infected in $\mathcal{M}$.
        This maps to the corresponding configuration~$\mathcal{H}\{M\}$ in the original network, where $s$ is estimated by summing the number of nodes in the recovered~($R$) state.
        \textbf{c2} Two conditions are required to preserve $s$ in the original network: 
        (1) $\beta \geq \hat{\beta_k}$ where $\hat{\beta_k}$ is the minimum transmission probability required for a seed node to infect all nodes within a $k$-clique~($G_{\Delta_k}$).
        (2) $\beta_{w_{ij}(k)} = 1-(1-\beta)^{e^w_{ij}(k)}$, where $\beta_{w_{ij}(k)}$ is the effective transmission probability for a weighted edge with weight $e^w_{ij}(k)$ in $G_{R=k}$.
              }
	\label{fig:fig1}
\end{figure*}

\label{subsec:ISCG-framework}
\subsection*{Iterative Structural Coarse-Graining Framework}
The iterative structural coarse-graining (ISCG) framework simplifies large-scale networks while preserving the contagion dynamics of the original network.
Achieving this requires addressing two critical challenges:
$(\romannumeral1)$ Ensuring that the total number of final infected nodes in the original network can be accurately reproduced by simulations on the coarse-grained networks (CGNs) using the same initial seeds.
$(\romannumeral2)$ Ensuring that the contagion process on the CGNs remains consistent with the dynamics of the original network.

To overcome these challenges,  ISCG iteratively merges maximal cliques---complete subgraphs where every pair of nodes is directly connected---into super-nodes.
These super-nodes are carefully designed to replicate the contagion dynamics of the original nodes they represent.
The dense connectivity within maximal cliques provides numerous transmission pathways, ensuring a high likelihood that if one node in a clique is infected, all nodes in the clique will eventually become infected.

In the coarse-grained representation, we hypothesize that all constituent nodes of a super-node in the original network are infected if the super-node itself becomes infected.
By summing the weights of infected super-nodes in a given contagion configuration $\mathcal{M}$, the total outbreak size in the original network can be accurately estimated~(Fig.~\ref{fig:fig1}\textbf{c1}).
This approach ensures that CGNs retain both macroscopic properties (e.g., total outbreak size) and microscopic properties (e.g., local infection dynamics), providing a scalable and reliable tool for analyzing large-scale networks.

\textbf{Iterative coarse-graining method}. 
We propose an iterative coarse-graining method to systematically simplify networks while preserving their dynamic properties.
Consider an undirected network $G(V, E)$, where $V$ and $E$ denote the sets of nodes and edges, respectively, and the network contains $N$ nodes.
Each node and edge in the network is initially assigned a unit weight, resulting in a weighted representation
$G(V_w, E_w)$, defined as follows:
\begin{equation}\label{eq:eq-1}
\begin{aligned}
V_w &=\left\{v^w_i|v^w_i=1,v_i\in V \right\}, \\
E_w &= \left\{e^w_{ij}|e^w_{ij} =1,e_{ij} \in E\right\}.
\end{aligned}
\end{equation}
The coarse-graining process is governed by an operator $\mathcal{F}$, which transforms the network $G$ into a coarse-grained network $G^{\prime} = \mathcal{F}(G)$.
At each iteration, maximal cliques in the network are identified and merged into super-nodes. 
When a node belongs to multiple maximal cliques, one clique is selected randomly for merging.
Connections between super-nodes, or between super-nodes and individual nodes, are represented as super-links.
The weights of super-nodes and super-links are updated to reflect the total number of original nodes and links they represent, calculated as the sum of the weights of their constituent nodes and links~(Fig.~\ref{fig:fig1}\textbf{a}).
This process is repeated iteratively until the network is reduced to a single node or a coarse-grained network (CGN) of the desired size.
During each iteration, all external links within a clique are assigned to the corresponding super-node, which may result in the formation of additional cliques in the CGN~(Fig.~\ref{fig:fig3}\textbf{b}).
This iterative method effectively reduces network size while retaining essential properties of the original network.

To formalize the process, we define a $k$-clique CGN, denoted as $G_{R=k}$, where the operator $\mathcal{F}$ has been applied iteratively until only cliques smaller than size $k$ remain:
\begin{equation}\label{eq:eq0}
   G_{R=k} = \mathcal{F}^{(t_k)}(G),  \quad 1 < k\leq \mathcal{K},  
\end{equation}
where $\mathcal{K}$ represents the size of the largest cliques in the original network $G$, and $t_k$ is the number of iterations required to reach $G_{R=k}$~(Fig.~\ref{fig:fig1}\textbf{b}). 
For instance, $G_{R=2}$, the $2$-clique CGN, reduces the network to a single node, with the node weight corresponding to the total number of nodes in the original network. 

\textbf{Conditions for preserving dynamic behavior}.
To ensure that the contagion dynamics on $k$-clique CGNs faithfully represent those of the original network, two key conditions must be fulfilled.
These conditions ensure the preservation of both macroscopic and microscopic properties of the contagion process across scales~(Fig.~\ref{fig:fig1}\textbf{c2}).
 
\textit{Condition 1}: Minimum transmission probability for complete contagion within cliques.
For $k$-clique CGNs to preserve the contagion dynamics of the original network, the transmission probability $\beta$ must exceed a threshold, $\hat{\beta}_k$, ensuring that a single infected seed triggers the full infection of all nodes within the $k$-clique.
This condition is given by:
\begin{equation}\label{eq:eq1}
    \beta \geq \hat{\beta}_k, \quad \hat{\beta}_k = \min{\mathop{\arg\max}\limits_{\beta} \Lambda^k_{1\rightarrow(k-1)}(\beta)}, 
\end{equation}
where $\Lambda^k_{1\rightarrow( k-1)}(\beta)$ represents the probability of a single infected seed successfully transmitting the infection to all other nodes in the $k$-clique.
A detailed derivation of $\Lambda^k_{1\rightarrow( k-1)}(\beta)$ is provided in the Eqs.~\ref{eq:eqs3} and~\ref{eq:eqs4}.
This condition guarantees that a super-node in the $k$-clique CGN accurately captures the complete infection dynamics of its constituent nodes in the original network~(Proposition 1).

\textit{Condition 2}: Effective transmission probability on weighted links.
To preserve the contagion dynamics between super-nodes in $k$-clique CGNs, the effective transmission probability $\beta_{w_{ij}(k)}$ along a weighted super-link must satisfy:
\begin{equation}\label{eq:eq2}
   \beta_{w_{ij}(k)} = 1-(1-\beta)^{e^w_{ij}(k)}. 
\end{equation}
where $e^w_{ij}(k)$ is the weight of the link between two super-nodes in the $k$-clique CGNs.
This condition ensures that the effective transmission  probability between super-nodes matches the cumulative transmission probability across all links connecting their corresponding components in the original network.
This relationship holds universally, no matter if the super-link connects individual nodes, subgraphs, or a combination of both~(Proposition 2).

Together, \textit{condition 1} guarantees the accurate representation of infection dynamics within each super-node, while \textit{condition 2} maintains consistent transmission probabilities across the reduced network.
These conditions form a rigorous foundation for retaining the essential dynamic properties of the original network in its coarse-grained representation. 
For detailed derivations and further implications, refer to the Methods section.

\begin{figure*}[!htbp]
	\centering
	\includegraphics[width=\textwidth]{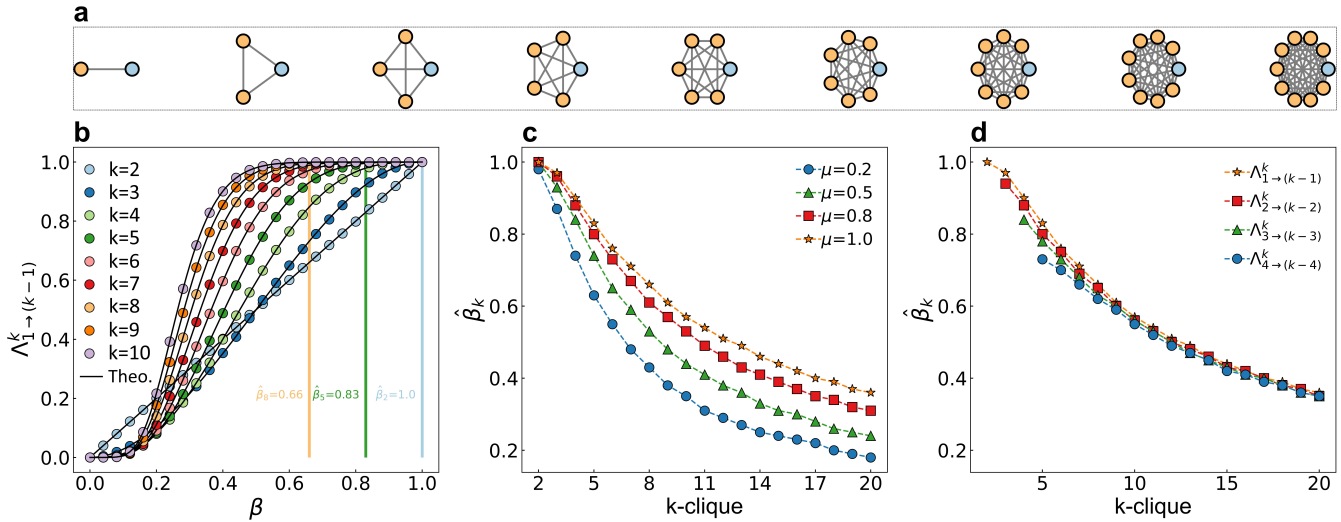}
	\caption{\textbf{Minimum transmission probability required to preserve the contagion behavior in $k$-clique CGNs.} 
	\textbf{a}  Schematic representation of a single seed node infecting all nodes within a $k$-clique~($k$ = 2 to 10). 
    Susceptible nodes are shown in orange, while the initially infected nodes are highlighted in blue.
	\textbf{b} Probability $\Lambda^k_{1\rightarrow(k-1)}$ 
    of a single seed node fully infecting a $k$-clique  as a function of the transmission probability $\beta$.
    Circle markers indicate numerical simulation results, while solid lines represent theoretical predictions obtained from Eqs~.\ref{eq:eqs3} and \ref{eq:eqs4}. 
    The vertical lines denote the minimum transmission probability $\hat{\beta}_k$, where $\Lambda^k_{1\rightarrow(k-1)}=1$.
	\textbf{c} Relationship between $\hat\beta_k$ and $k$-clique size for different recovery probabilities $\mu$. 
    Larger cliques require smaller $\hat\beta_k$ for full contagion.
	\textbf{d} Influence of multiple initial seed nodes on $\hat{\beta}_k$ for varying $k$-clique sizes with $\mu=1$.
    The difference in $\hat{\beta}_k$ between single-seed and multi-seed scenarios diminishes as clique size increases.}
	\label{fig:fig2}
\end{figure*}

\subsection*{Complete contagion in \texorpdfstring{$k$}{k}-cliques}
The minimum transmission probability $\hat{\beta}_k$ is a fundamental parameter that ensures complete contagion within $k$-cliques, directly influencing the applicability of the ISCG framework.
To derive $\hat{\beta}_k$, we compute the probability $\Lambda^k_{1\rightarrow(k-1)}$ of all nodes in a $k$-clique becoming infected by a single seed node and set it to 1.
For simplicity, we adopt the SIR model with $\mu=1$, equivalent to a bond percolation process.

In a symmetric $k$-clique ($k\geq2$, Fig.~\ref{fig:fig2}\textbf{a}), the probability of  full contagion starting from a single seed node is expressed as: 
\begin{equation}\label{eq:eqs3}
	\Lambda^k_{1\rightarrow (k-1)}(\beta) = \sum^{k-1}_{m=1}\tbinom{k-1}{m}\beta^m(1-\beta)^{k-1-m} \Lambda^{k-1}_{m\rightarrow(k-1-m)}(\beta),
\end{equation}
where $m$ is the number of nodes initially infected by the seed, and $\Lambda^{k-1}_{m\rightarrow(k-1-m)}(\beta)$ accounts for the subsequent infection of remaining susceptible nodes.
Expanding this term further, we enumerate all non-infection events and derive:
\begin{equation}\label{eq:eqs4}
\begin{aligned}
	\Lambda^{k-1}_{m\rightarrow(k-1-m)}(\beta) &= 1-\sum_{n=0}^{k-2-m}\tbinom{k-1-m}{n}(1-\beta)^{m(k-1-m-n)}\\
    & \quad \cdot (1-\beta)^{n(k-1-m-n)}\Lambda^{m+n}_{m\rightarrow n}(\beta),
\end{aligned}
\end{equation}
where the summation enumerates all events in which contagion fails to spread further.
It is important to note that $\Lambda^m_{m\rightarrow0}(\beta)=1$ for any $m$ and $\beta$, as this corresponds to a scenario where no further infections occur.
To illustrate the iterative nature of this relationship,  consider the case $k=Z$ and $m=1$ in Eq.~\ref{eq:eqs4}, we obtain 
\begin{equation}
\Lambda^{Z-1}_{1 \rightarrow (Z-2)}(\beta) = \Lambda^{(Z-1)}_{1 \rightarrow [(Z-1)-1]}(\beta), 
\end{equation}
where the right-hand side is equivalent to Eq.~\ref{eq:eqs3} with $k=Z-1$.
This recursive relationship highlights the iterative structure of Eqs.~\ref{eq:eqs3} and~\ref{eq:eqs4}, allowing for the exact computation of 
$\Lambda^k_{1\rightarrow (k-1)}(\beta)$ without relying on extensive numerical simulations.
By solving these equations iteratively, we can determine the minimum transmission probability $\hat{\beta}_k$ required for complete contagion within a $k$-clique.
For a detailed derivation, including recursive formulas, example calculations for small cliques, and extensions to larger cliques, see \SIM: Recursive calculation of full infection probability in $k$-cliques.

\begin{figure*}[!htbp]
	\centering
	\includegraphics[width=\textwidth]{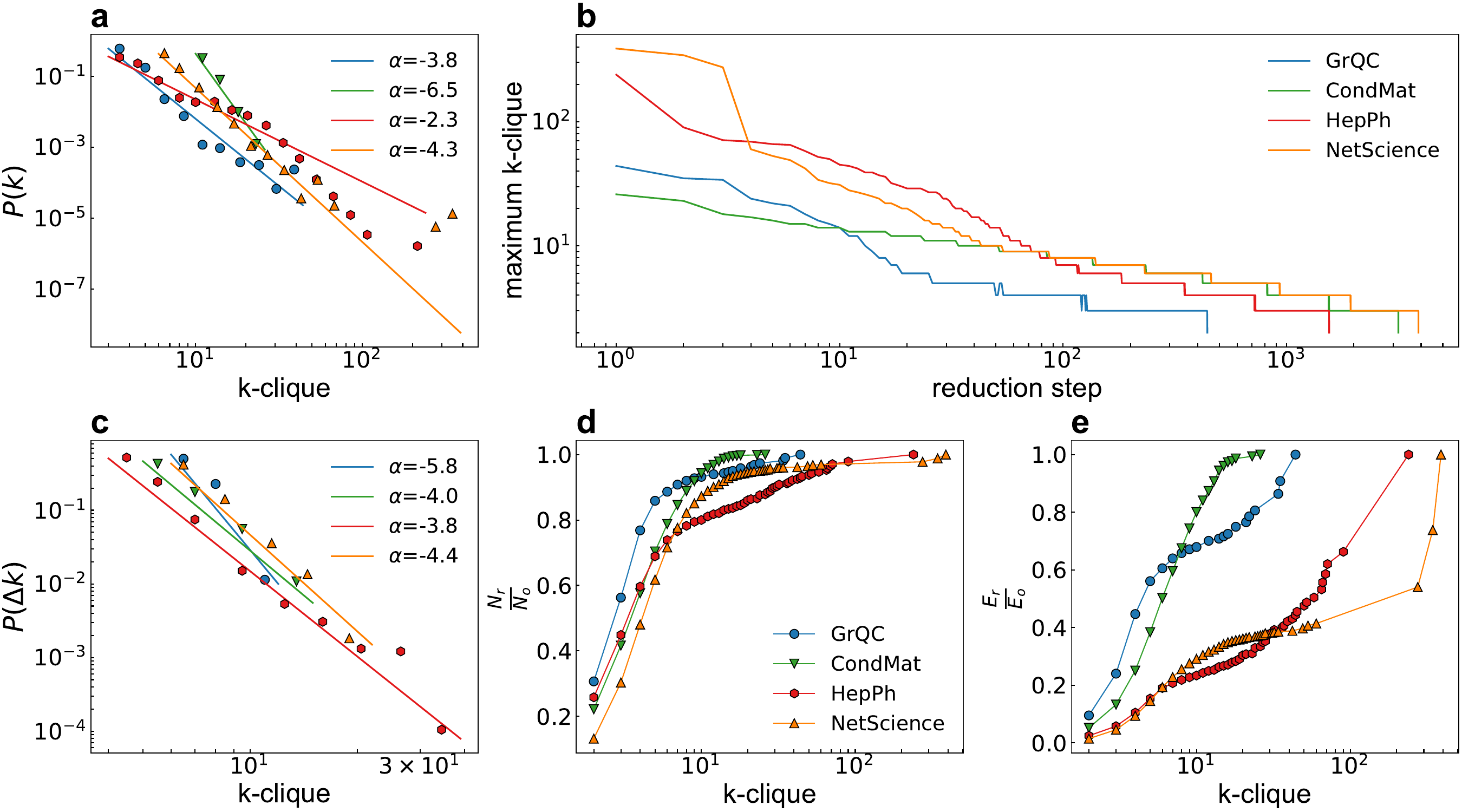}
	\caption{\textbf{Network reduction performance of $k$-clique CGNs on four real-world networks.} 
	\textbf{a} Distribution of clique size in the original networks, following a power-law relationship $P(k)\sim k^{-\alpha}$, where $\alpha$ is fitted using maximum likelihood estimation. 
	\textbf{b} Size of the largest clique as a function of iteration steps during the reduction process.
    The curves reveal that merging smaller cliques often generates new $k$-cliques, extending the reduction process as $k$ decreases.
	\textbf{c} Distribution of the size of newly formed cliques during the coarse-graining process (excluding those present in the original network). 
    The size of these newly formed cliques also follow a power-law distribution, consistent with the original network.
	\textbf{d} Proportion of nodes remaining in $k$-clique CGNs relative to the original network.
    The proportion decreases sharply as $k$ decreases, particularly for smaller $k$. 
	\textbf{e}  Proportion of links remaining in $k$-clique CGNs relative to the original network. 
    The number of links decreases exponentially, resulting in substantial network simplification.
    } 
	\label{fig:fig3}
\end{figure*}

We validate these analytical results by comparing $\Lambda^k_{1\rightarrow(k-1)}$ against numerical simulations for cliques of varying sizes~(Fig.~\ref{fig:fig2}\textbf{b}).
The excellent agreement between theoretical and simulation results confirms the accuracy of our derivations.
The identification of $\hat{\beta}_k$ provides several key insights into the behavior of contagion dynamics within $k$-clique.
First, as shown in Fig.~\ref{fig:fig2}\textbf{b}, the minimum transmission probability $\hat{\beta}_k$ decreases as the clique size $k$ increases.
This implies that if full contagion occurs in a $k$-clique at $\hat{\beta}_k$, it will also occur in any larger clique. 
Second, we observe a trade-off between network reduction and dynamic fidelity.
Reducing the network to smaller cliques imposes stricter conditions~(higher $\hat{\beta}_k$) to maintain the contagion dynamics.
In contrast, when the network is reduced to higher-order $k$-clique CGNs, the constraints on transmission probability are relaxed, making it easier to preserve contagion dynamics~(Fig.~\ref{fig:fig2}\textbf{c}).
Third, in cases where contagion is initiated by multiple seed nodes within a clique, the required transmission probability $\hat{\beta}_k$ decreases compared to the single-seed scenario~(Fig.~\ref{fig:fig2}\textbf{d}).
However, as the clique size increases, this difference diminishes.
This indicates that the influence of multiple seeds becomes negligible in larger cliques.
Notably, this also suggests that in higher-order $k$-clique CGNs, the contagion behavior of a super-node being infected by multiple external nodes more closely matches that of the original network compared to lower-order CGNs.
The derivation and validation of $\hat{\beta_k}$ provide a quantitative foundation for ensuring complete contagion within $k$-cliques.
These findings underscore the balance between network reduction and dynamic fidelity, providing a deeper understanding of the ISCG framework’s ability to preserve contagion dynamics across different scenarios.

\section{Result}
\label{sec:results}

\begin{figure*}[!htbp]
	\centering
	\includegraphics[width=\textwidth]{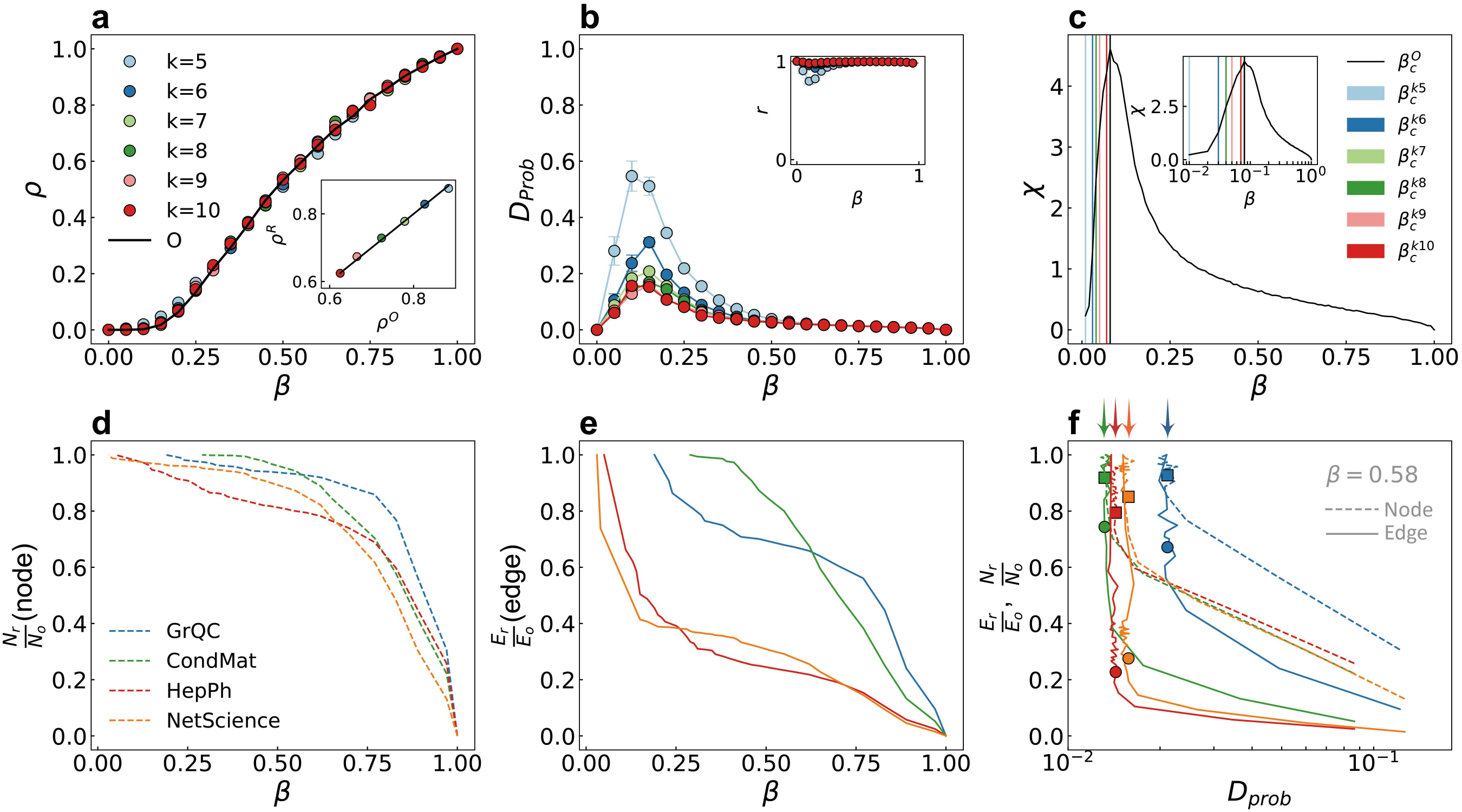}
	\caption{\textbf{Preservation of SIR contagion dynamics on $k$-clique CGNs compared to the original network}.
    \textbf{a} Final density of infected nodes as a function of the transmission probability $\beta$ on $k$-clique CGNs~($k$ = 5 to 10, circle markers) and the original network~(solid line).
    Results closely match across all scales.
    Inset: Comparison of the final infected nodes density~($\rho^R$) on $k$-clique CGNs and the original networks~($\rho^O$) at the threshold of transmission probability $\beta = \hat{\beta}_k$. 
	\textbf{b} Distance $D_{prob}$ between the infection probability vector of the original network ($v_O$) and that of $k$-clique CGNs~($v_R$) as a function of $\beta$, where each element in vector represents the probability of a node being infected. 
	Here, $D_{prob} = (\sum_{i=1}^N(v_O^i-v_R^i)^2)^{1/2}$, with vectors normalized to unit length. 
	Smaller $D_{prob}$ indicates higher accuracy in preserving the node-level contagion probabilities. 
	Inset: Pearson correlation coefficient $r$ between the two vectors, illustrating the degree of alignment.
	\textbf{c} Susceptibility $\chi$ as a function of $\beta$, where $\chi = [\langle s^2 \rangle -\langle s \rangle ^2]/\langle s \rangle$, with $s$ representing the outbreak size in a single simulation. 
	Susceptibility measures the fluctuation in outbreak sizes, reaching its peak near critical transitions~\cite{radicchi2015predicting}.
    This peak corresponds to the critical transmission probability $\beta_c$, which is identified for both the original network~(black) and $k$-clique CGNs~(colored lines).
    Inset: Log-scale plot of $\beta$ to emphasize detailed behavior near critical points. 
	\textbf{d} Proportion of nodes in the smallest-scale $k$-clique CGNs required to preserve contagion dynamic for a given $\beta$~(dashed line), emphasizing the reduction in node number under exact reduction conditions.
    \textbf{e} Proportion of edges in the smallest-scale $k$-clique CGNs under the same condition~(solid line), showcasing the effective reduction of edges while ensuring dynamic fidelity.
	\textbf{f} At $\beta=0.58$, the proportion of nodes~(dashed) and edges~(solid) are plotted against $D_{prob}$ for different $k$-clique CGNs.  
    Benchmark values for the 10-clique CGNs, marked by squares and circles, represent exact reduction conditions, ensuring full preservation of contagion dynamics at $\beta=\hat{\beta}_{10}=0.58$~(see Fig.~\ref{fig:fig2}c). 
    Arrows highlight the corresponding $D_{prob}$ values, providing a reference for the accuracy in maintaining dynamic behaviors.
    With further reductions beyond these baselines, $D_{prob}$ increases; however, the networks still achieve substantial complexity reduction with only minimal precision loss, underscoring the ISCG framework's effectiveness under approximate reduction.
    For further details on numerical simulations, see the Methods section.
    Results in \textbf{a-c} are obtained from the GrQC network. 
    }
	\label{fig:fig4}
\end{figure*}

\subsection*{Reduction ratio of  \texorpdfstring{$k$}{k}-clique CGNs}
Effectively reducing network complexity is essential for analyzing large-scale systems.
Here, we evaluate the performance of the ISCG framework in simplifying networks through iterative coarse-graining. 
To demonstrate its applicability, we analyze four real-world collaborative networks~\cite{Ryan2015}---GrQc, CondMat, HepPh, and NetScience---representing diverse scientific communities.
Additional results for other network types, including their structural characteristics and reduction outcomes, are provided in the \SI{} and \SIR{}.

Figure~\ref{fig:fig3}\textbf{a} shows the distribution of clique sizes in the original networks, which follows a power-law relationship with steep exponents~(except for HepPh). 
This observation indicates that most cliques are relatively small, suggesting fewer iterations are needed to merge them during the coarse-graining process. 
However, Figure~\ref{fig:fig3}\textbf{b} reveals a more complex behavior: the reduction process exhibits extended plateaus, where merging small cliques leads to the formation of new, predominantly smaller cliques.
This dynamic is further illustrated in Fig.~\ref{fig:fig3}\textbf{c}, where the distribution of
newly generated cliques during the coarse-graining process also exhibits a power-law form. 
This behavior highlights the adaptability of the ISCG framework in preserving fine-grained structural details across multiple scales, even as the network is progressively simplified.

To quantify the efficiency of the reduction process, 
Figures~\ref{fig:fig3}\textbf{d} and \textbf{e} show the proportions of remaining nodes~($N_r/N_o$) and edges ($E_r/E_0$) in $k$-clique CGNs relative to the original network.
Both metrics decrease rapidly as $k$ decreases, underscoring the framework's ability to achieve substantial simplification.
Notably, at the 5-clique level, the number of edges is reduced by approximately 50\% across most networks.
Larger networks, such as NetScience~(details provided in \SI), exhibit particularly pronounced reductions, demonstrating the ISCG framework’s scalability and efficiency in handling large and complex systems.
The ISCG framework’s versatility extends beyond collaborative networks, as it has been successfully applied to various real-world systems, including social, communication, and citation networks (Supplementary Figure 3). 
These results consistently confirm the framework’s robustness in achieving significant size reductions while retaining essential properties.
This dual capability enhances computational efficiency and reduces storage requirements, providing a practical advantage for studying large-scale dynamic processes.

Additionally, the ISCG framework offers significant flexibility through its approximate reduction capabilities (Supplementary Figure 4). 
By relaxing the strict structural constraints of $k$-cliques and employing $k$-plexes---substructures that allow up to $k$ missing edges---the framework maintains comparable reduction performance while significantly improving computational scalability. 
This relaxation enables the framework to handle diverse and less dense topologies more effectively, which is particularly beneficial for large-scale systems where computational costs are prohibitive.
The ISCG framework’s ability to balance network reduction and computational scalability makes it a powerful and versatile tool for coarse-graining networks across a wide range of applications.

\subsection*{Preservation of contagion dynamics on \texorpdfstring{$k$}{k}-clique CGNs}
We evaluate the ISCG framework’s ability to preserve contagion dynamics across coarse-grained networks at various levels of reduction. 
Using $k$-clique CGNs, we assess its performance through extensive simulations, focusing on macroscopic and microscopic behaviors, as well as critical contagion thresholds.
The results demonstrate that ISCG achieves significant network simplification while maintaining dynamic fidelity with high accuracy, offering a scalable tool for analyzing contagion processes.

Macroscopic preservation: final outbreak size.
Figure~\ref{fig:fig4}\textbf{a} shows that the density of final infected nodes on $k$-clique CGNs~($k=5$ to $10$)  closely aligns with results from the original network across all transmission probabilities $\beta$.
At the threshold of transmission probability $\hat{\beta}_k$,
where full contagion occurs within each $k$-clique, the values from $k$-clique CGNs are in perfect agreement with those of the original network (Figure~\ref{fig:fig4}\textbf{a}, inset).
This consistency validates the framework’s ability to accurately preserve outbreak sizes.
At the microscopic scale, the ISCG framework retains high accuracy in capturing node-level contagion probabilities. 
As shown in Figure~\ref{fig:fig4}\textbf{b}, the Euclidean distance $D_{prob}$ which measures the discrepancy between infection probability vectors from the original network and $k$-clique CGNs, decreases towards zero as $\beta$ increases, with near-perfect alignment observed around $\hat{\beta}_{10} = 0.58$.
Even at lower transmission probabilities~($\beta \ll \hat{\beta}_k$), the Pearson correlation between the probability vectors remains high (Figure~\ref{fig:fig4}\textbf{b}, inset), demonstrating the robustness of ISCG in preserving node-level dynamics across coarse-grained representations.
In addition, the ISCG framework also accurately captures critical contagion behavior. 
Figure~\ref{fig:fig4}\textbf{c} shows that the critical transmission probability $\beta_c$, identified by peaks in susceptibility $\chi$, remains close between $k$-clique CGNs and the original network, even when $\beta_c$ is significantly below $\hat{\beta}_k$. 
This result highlights the framework’s ability to maintain critical thresholds across various scales of network reduction.
To further validate the framework’s robustness, we extended our analysis to a range of network types, assessing its ability to preserve both macroscopic and microscopic contagion dynamics as well as critical thresholds.
The results, presented in Supplementary Figure~\ref{fig:figS6}, demonstrate that ISCG consistently achieves high accuracy in maintaining contagion processes across different topologies and scales.

As demonstrated previously, the ISCG framework ensures accurate preservation of contagion dynamics on coarse-grained networks when the condition $\beta \geq \hat{\beta_k}$ is satisfied.
This finding establishes a theoretical minimum reduction scale necessary to reliably maintain the original network's contagion behavior for a given $\beta$. 
Building on this, we analyze the relationship between reduction ratios and $\beta$, focusing on the extent to which nodes and edges can be reduced while precisely preserving contagion dynamics.
Figures~\ref{fig:fig4}\textbf{d} and \textbf{e} depict the reduction ratios for nodes ($N_r / N_o$) and edges ($E_r / E_o$) as a function of $\beta$. 
Node reduction is most pronounced at higher $\beta$, 
where contagion spreads more readily, while edge reduction remains significant across a wide range of $\beta$ values.
For instance, at lower $\beta$, the HepPh and NetScience networks achieve a 50\% edge reduction, underscoring ISCG’s scalability and computational efficiency.
These results demonstrate the framework’s capacity to provide substantial simplifications while maintaining the fidelity of contagion dynamics, making a robust tool for analyzing large-scale networks under diverse conditions.

In addition to the exact reductions shown in Figs.~\ref{fig:fig4}\textbf{d} and \textbf{e}, we also examine approximate reductions by relaxing the condition $\beta \geq \hat{\beta_k}$.
This analysis explores how further reductions in network size affect the accuracy of microscopic dynamics.
Figure~\ref{fig:fig4}\textbf{f} illustrates the trade-off between reduction ratios ($N_r / N_o$ and $E_r / E_o$) and accuracy measured as $D_{prob}$ at a fixed $\beta = 0.58$.
The markers and arrows highlight the minimum reduction scale required to fully preserve contagion dynamics under the exact condition $\beta = \hat{\beta}_{10}$, providing a benchmark for comparison. 
Up to this threshold~($\beta = \hat{\beta}_{10}$), $D_{prob}$ remains unchanged despite significant reductions in nodes and edges, demonstrating the ISCG framework’s ability to precisely maintain contagion dynamics at the theoretical threshold. 
Beyond the threshold, as further reductions are applied, $D_{prob}$ increases due to the trade-off between simplicity and fidelity. 
However, networks such as CondMat, HepPh, and NetScience notably achieve substantial reductions with only minimal precision loss.
This behavior persists across a range of transmission probabilities $\beta$ (e.g, $\beta = 0.36$ and $\beta = 0.83$, Supplementary Figure 7), underscoring the framework’s adaptability and robustness.
This finding highlights that the ISCG framework not only achieves significant complexity reduction but also allows for controlled accuracy loss under relaxed conditions, offering flexibility to meet diverse practical requirements.

\begin{figure*}[!htbp]
	\centering
	\includegraphics[width=\textwidth]{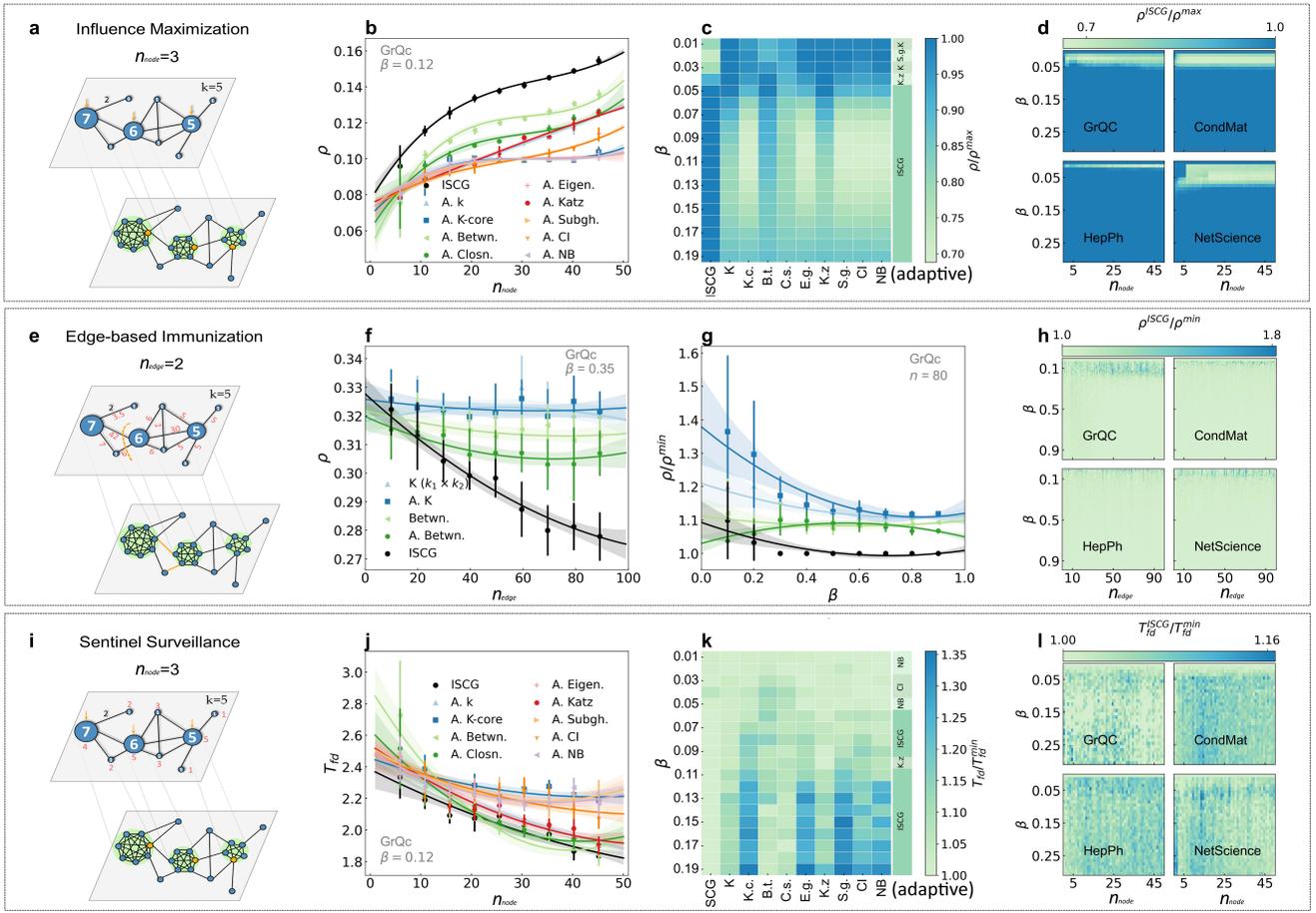}
		\caption{\textbf{Application of ISCG-based method to three classical contagion problems}. 
        The ISCG framework leverages $k$-clique CGNs to reduce network complexity and efficiently address three key contagion challenges: influence maximization, edge-based immunization, and sentinel surveillance.
        By leveraging coarse-grained representations, these methods balance computational efficiency with high performance in identifying optimal nodes or edges.
        \textbf{a-d Influence maximization}.
        \textbf{a} ISCG-IM($k=5$) ranks super-nodes in the 5-clique CGN based on their weights to identify the top $n_{node}$ candidate seed sets (orange arrows). 
        From each candidate set (highlighted in green), the node with the highest degree in the original network is selected to form the final seed set (orange nodes).
        \textbf{b} The performance of ISCG-IM as a function of the number of seeds ($n_{node}$) at $\beta = 2\beta_c$.
        \textbf{c} Comparison of influence maximization methods, evaluated by maximal average outbreak size ($\rho^{max}$) as a function of transmission probability ($\beta$) with $n_{node}=30$.
        \textbf{d} Heatmap of ISCG-IM performance across four networks as a function of $\beta$ and $n_{node}$.
        \textbf{e-h Edge-based immunization}.
        \textbf{e}  ISCG-EI ($k=5$) identifies critical edges by scoring super-edges in the 5-clique CGN. Scores~(red numbers) are computed as the product of the weights of connected super-nodes divided by the super-edge weight. 
        The algorithm iteratively selects the highest-scoring super-edges, calculates their minimum edge cuts~(orange dashed lines), maps these cuts back to the original network, and removes the corresponding $n_{edge}$ edges~(orange edges).
        \textbf{f} Performance of ISCG-EI as a function of the number of removed edges ($n_{edge}$) at $\beta = 5\beta_c$.
        \textbf{g} Comparison of immunization methods, evaluated by minimal average outbreak size ($\rho^{min}$) as a function of $\beta$ with $n_{edge}=80$.
        \textbf{h} Heatmap of ISCG-EI performance across four networks as a function of $\beta$ and $n_{edge}$.
        \textbf{i-l Sentinel surveillance}.
        \textbf{i} ISCG-SS ($k=5$) scores super-nodes in the 5-clique CGN based on the sum of weights of connected super-edges~(red numbers) to identify the top $n_{node}$ candidate sentinel sets~(orange arrows). From each candidate set, the node with the highest degree in the original network is chosen to form the final sentinel set~(orange nodes).
        \textbf{j} Performance of ISCG-SS, measured by the earliest time to detect an outbreak ($T_{fd}^{min}$) as a function of $n_{node}$ at $\beta = 2\beta_c$.
        \textbf{k} Comparison of sentinel surveillance methods, evaluated by $T_{fd}^{min}$ as a function of $\beta$ with $n_{node}=30$.
        \textbf{l} Heatmap of ISCG-SS performance across four networks as a function of $\beta$ and $n_{node}$.
        \textbf{c, j, k} The bar on the right of each panel highlights the best-performing method for each $\beta$. 
        Results in \textbf{c, k} are from the GrQC network; other panels specify their respective networks.
        }
	\label{fig:fig5}
\end{figure*}

To further enhance computational efficiency, we apply the ISCG framework to unweighted CGNs, disregarding edge weights to focus solely on structural contributions to contagion dynamics.
This approach significantly reduces computational overhead while simplifying network representations. 
Despite the omission of edge weights, unweighted CGNs maintain strong performance in preserving both macroscopic and microscopic contagion behaviors, with only minor accuracy trade-offs compared to their weighted counterparts.
Detailed comparisons with weighted CGNs (Supplementary Figures~\ref{fig:figS6} and \ref{fig:figS7}) underscore the trade-off between simplicity and accuracy, highlighting the critical role of edge weights in capturing finer-grained dynamics.
These results demonstrate the flexibility of the ISCG framework, making it well-suited for applications that require simplified yet effective network representations.

Overall, the ISCG framework provides a robust solution for balancing network reduction with dynamic fidelity. 
It preserves contagion dynamics across scales, achieves substantial complexity reductions, and offers adaptability for both exact and approximate representations. 
Its scalability and versatility make ISCG an invaluable tool for analyzing contagion processes in large-scale networks across diverse scenarios.

\subsection*{Application of ISCG framework}
The ISCG framework leverages $k$-clique CGNs to simplify complex networks by aggregating localized dense connections into super-nodes while preserving essential contagion dynamics. 
This approach reduces computational complexity, highlights key structural components driving network behavior, and provides a foundation for solving practical contagion-related challenges.
To demonstrate its utility, we apply the ISCG framework to three classical contagion problems~\cite{holme2017three}: influence maximization (IM), edge-based immunization (EI), and sentinel surveillance (SS). 
For each problem, tailored strategies---ISCG-IM, ISCG-EI, and ISCG-SS---are developed to identify critical nodes, edges, or structural elements using coarse-grained representations.
Extensive comparisons reveal that ISCG-based methods outperform traditional adaptive centrality approaches, showcasing their scalability, robustness, and practical utility.

\textbf{Influence maximization}.
Influence maximization seeks to identify seed nodes that maximize the spread of influence, typically quantified as the average density of infected nodes~($\rho$)~\cite{kitsak2010, chen2009efficient, morone2015influence, xue2022}.
The ISCG-IM method addresses this challenge by leveraging the coarse-grained structure of $k$-clique CGNs to estimate influence potential at both macroscopic and microscopic scales.
This dual-scale approach not only enhances computational efficiency but also improves accuracy in identifying influential nodes.

The ISCG-IM method combines macroscopic insights from $k$-clique CGNs with microscopic node-level information to optimize seed selection. 
This dual-scale approach identifies super-nodes as critical structural components and pinpoints the most influential nodes within them. 
The method proceeds as follows:
$\mathrm{(i)}$ Candidate seed selection:
Super-nodes in the $k$-clique CGN are ranked based on their weights, which reflect the aggregated influence of their constituent nodes.
The top $n_{node}$ super-nodes, corresponding to the most influential macroscopic components, are selected as candidate seed sets.
$\mathrm{(ii)}$ Final seed selection:
From each candidate set, the node with the highest degree in the original network is chosen as the final seed, ensuring effective propagation within the corresponding super-node~(Figure~\ref{fig:fig5}\textbf{a}).
This strategy enhances influence coverage by distributing seeds across distinct super-nodes, avoiding diminishing returns in outbreak size ($\rho$) that occur when multiple seeds are concentrated within the same super-node~(Supplementary Figures~\ref{fig:figS8}\textbf{a} and \textbf{b}).

The performance of ISCG-IM was benchmarked against a wide range of adaptive centrality-based methods, including degree($k$), k-core~\cite{kitsak2010}, betweenness~\cite{freeman2002centrality}, closeness~\cite{sabidussi1966centrality}, eigenvector~\cite{bonacich2007some}, Katz~\cite{zhan2017identification}, subgraph~\cite{estrada2005subgraph}, collective influence~\cite{morone2015influence}, and non-backtracking centrality~\cite{martin2014localization}.
These methods iteratively recalculate node importance after each selection,  dynamically adapting to structural changes in the network. 
While this provides a strong baseline, ISCG-IM consistently outperforms these approaches across multiple metrics.
As shown in Fig.~\ref{fig:fig5}\textbf{b}, ISCG-IM achieves significantly larger outbreak sizes across various seed numbers ($n_{node}$) in the GrQC network at $\beta = 2\beta_c$.
Figure~\ref{fig:fig5}\textbf{c} highlights its robust performance across a range of transmission probability ($\beta$), achieving higher maximal outbreak sizes at larger $\beta$ values.
Scalability and generalizability are evident in Fig.~\ref{fig:fig5}\textbf{d}, where the ratio of the final outbreak size achieved by ISCG-IM ($\rho^{ISCG}$) to the maximum average outbreak size observed from all methods demonstrates strong performance across diverse networks (GrQC, CondMat, HepPh, NetScience) under varying $\beta$ and $n_{node}$.
Additional insights into ISCG-IM’s mechanisms are provided in Supplementary Figure~\ref{fig:figS8}, illustrating its ability to integrate macroscopic and microscopic features for optimal seed selection.

\textbf{Edge-based immunization}.
Edge-based immunization targets critical contagion pathways to curtail the spread of infection~\cite{schneider2011suppressing, chen2008finding}.
ISCG-EI leverages $k$-clique CGNs to simplify the identification and removal of key edges that maintain network functionality.
The ISCG-EI approach comprises two key steps:
$\mathrm{(i)}$ Edge Scoring: Super-edges in the $k$-clique CGN are scored based on the product of the weights of connected super-nodes divided by the super-edge weight~(Fig.~\ref{fig:fig5}\textbf{e}, red numbers).
This score quantifies an edge’s contribution to contagion spread.
$\mathrm{(ii)}$ Critical edge identification: High-scoring super-edges are iteratively selected, and their corresponding minimum edge cuts in the original network are identified and removed~(Fig.~\ref{fig:fig5}\textbf{e}, orange edges).

As shown in Fig.~\ref{fig:fig5}\textbf{f}, ISCG-EI shows lower average infection densities ($\rho$) as the number of removed edges ($n_{edge}$) increases.
Figure~\ref{fig:fig5}\textbf{g} further demonstrates ISCG-EI’s robustness across a range of transmission probabilities $\beta$, achieving the lowest ratio $\rho/\rho^{min}$ of average infection density to the minimum observed value ($\rho^{min}$) among all methods with a given $\beta$. 
Scalability is illustrated in Fig.~\ref{fig:fig5}\textbf{h}, where ISCG-EI consistently outperforms benchmarks across diverse real-world networks.
Supplementary Figure~\ref{fig:figS9} provides additional analysis, emphasizing ISCG-EI’s ability to effectively prioritize and disrupt critical contagion pathways.

\textbf{Sentinel Surveillance}.
Sentinel surveillance identifies nodes most likely to detect contagion early, enabling timely intervention~\cite{colman2019efficient, herrera2016disease}.
ISCG-SS combines macroscopic insights with microscopic refinements to select optimal sentinel nodes efficiently.
$\mathrm{(i)}$ Candidate seed selection: 
Super-nodes in the $k$-clique CGN are scored based on the sum of the weights of connected super-edges (Fig.~\ref{fig:fig5}\textbf{i}, red numbers).
High scores prioritize nodes likely to encounter contagion early.
$\mathrm{(ii)}$  From each candidate set, the node with the highest degree in the original network is chosen, ensuring effective detection at a granular level~(Fig.~\ref{fig:fig5}\textbf{i}, orange nodes).

Performance evaluation in Fig.~\ref{fig:fig5}\textbf{j} demonstrates that ISCG-SS achieves significantly earlier detection times ($T_{fd}$) compared to competing methods in the GrQC network at $\beta = 2\beta_c$.
Figure~\ref{fig:fig5}\textbf{k} highlights its robustness across varying transmission probabilities $\beta$.
Scalability is evident in Fig.~\ref{fig:fig5}\textbf{l}, where ISCG-SS maintains strong performance across diverse  real-world networks.
Supplementary Figure~\ref{fig:figS10} provides detailed insights into the underlying mechanisms of ISCG-SS, showing how it optimally integrates macroscopic and microscopic insights to prioritize sentinel nodes.

By leveraging coarse-grained representations, these results demonstrate the versatility and effectiveness of ISCG-based methods in addressing diverse contagion challenges.
More importantly, beyond the specific performance of the current method, they offer novel insights into uncovering critical structural elements hidden within large-scale networks, opening new avenues for designing more efficient methods to tackle a wide range of real-world contagion problems.

\section{Discussion}
In this paper, we introduce the iterative structural coarse-graining~(ISCG) framework, a novel methodology designed to simplify complex networks and preserve their essential contagion dynamics.
By leveraging $k$-clique coarse-grained networks~(CGNs), ISCG effectively reduces the computational complexity of analyzing large-scale networks, maintaining high fidelity across multiple scales of contagion processes.

Our theoretical analyses and numerical simulations on diverse datasets demonstrate that $k$-clique CGNs accurately replicate the contagion dynamics of the original network under the derived condition $\beta \geq \hat{\beta}_k$.
Specifically, ISCG preserves key dynamical properties, including final outbreak size, node-level infection probabilities, and phase transition behaviors. 
These results confirm ISCG’s robustness in maintaining both macroscopic and microscopic dynamics, ensuring high fidelity in contagion modeling even after substantial network reductions.
Furthermore, our exploration of both exact and approximate reductions reveals that ISCG achieves substantial complexity reduction without compromising critical dynamical properties.
This balance between simplification and fidelity highlights ISCG’s potential as a scalable and practical tool for studying contagion processes in complex networks.

Building on the multi-scale representations provided by $k$-clique CGNs, ISCG addresses real-world challenges such as influence maximization, edge-based immunization, and sentinel surveillance. 
The framework uncovers critical structural elements, such as influential nodes and essential edges, which drive contagion dynamics but are often obscured in large-scale networks. 
By revealing these hidden patterns, ISCG not only enhances computational efficiency but also provides actionable insights for tackling challenges in epidemic control, information dissemination, and infrastructure resilience. 
These results emphasize ISCG’s dual role as both an analytical tool and a practical framework for solving contagion-related problems.

Beyond its immediate applications, the ISCG framework has far-reaching implications for elucidating the intricate interplay between structure and dynamics. 
Unlike conventional coarse-graining methods that focus on pairwise interactions, ISCG uniquely captures higher-order structural patterns, such as simplicial complexes and hypergraphs, which are increasingly recognized as essential for modeling complex systems. 
By preserving the integrity of higher-order interactions during network reduction, ISCG provides a deeper insights into how interactions beyond simple pairwise relationships drive key dynamical processes~\cite{iacopini2019simplicial, alvarez2021evolutionary, battiston2021physics, zhang2023higher, ferraz2024contagion}.
For instance, in complex contagion dynamics, ISCG uncovers how densely connected groups amplify spreading through synergistic effects, accelerating transmission across the network~\cite{iacopini2019simplicial, de2020social, battiston2021physics}.
It also identifies critical groups that act as amplifiers or bottlenecks, shaping contagion pathways and boundaries.

Furthermore, the ISCG framework opens new pathways for understanding complex systems by analyzing hierarchical structures, self-organization, and emergent behaviors. 
It facilitates the identification of overlapping communities and multi-scale functional units~\cite{palla2005uncovering, ahn2010link, mucha2010community}, moving beyond traditional modularity-based approaches.
This capability is particularly valuable in social, biological, and communication networks, where clustered and overlapping structures often govern collective behaviors~~\cite{alvarez2021evolutionary, battiston2021physics, zhang2023higher}.
In addition to its structural insights, ISCG  naturally applies to a broad range of contagion processes, including reversible epidemics~(e.g., SIS), multi-state contagion processes~(e.g., SEIR), and complex contagion models~\cite{pastor2015, iacopini2019simplicial, de2020social, ferraz2024contagion}. 
Beyond contagion, ISCG extends to diverse phenomena such as synchronization in neural and power-grid networks, cascading failures in critical infrastructure, and opinion dynamics in social systems~\cite{alvarez2021evolutionary, battiston2021physics, gross2020two, zhang2023higher}.
By integrating structural and dynamical perspectives, ISCG transcends its role as a computational tool, serving as a conceptual framework for addressing complex challenges across diverse domains.

While the ISCG framework provides significant insights across various aspects,  it also has limitations that present opportunities for further research. 
A fundamental principle of the current framework is that complete transmission among constituent nodes occurs within supernodes (i.e., aggregated higher-order structures).
This principle ensures precise retention of contagion pathways during network reduction but may require further refinement to address scenarios involving partial or incomplete transmission, such as contagion processes with extremely low transmission probabilities (see the Accuracy Analysis of ISCG Framework for details in \SIM{}).
Hence, developing a more general coarse-graining approach that accounts for non-full contagion within supernodes could significantly enhance the framework’s applicability to real-world systems.
Additionally, the framework’s utility in non-contagion dynamics, such as opinion formation or cascading failures, remains largely unexplored. 
Rigorous evaluation of its performance in these contexts, along with domain-specific modifications, could broaden its applicability. 

\section{Methods}
\subsection*{Consistency of contagion dynamics}
Preserving contagion dynamics during the coarse graining process is fundamental to the ISCG framework.
Ensuring that the simplified representation replicates the dynamic behaviors of the original network is essential for achieving computational efficiency without sacrificing analytical accuracy.
In this section, we establish the theoretical foundation for this consistency.
First, we define the inverse operators that map the CGN to its corresponding original network.
Using these definitions, we derive two key conditions for maintaining contagion dynamics: the equivalence of final outbreak size and the equivalence of contagion configuration probabilities.
These conditions are rigorously validated through theoretical proofs in Propositions 1 and 2, demonstrating the robustness and accuracy of the ISCG framework.
 
\textbf{Inverse coarse-grained operator $\mathcal{H}\{*\}$}. 
The inverse coarse-grained operator $\mathcal{H}\{*\}$ formalizes the relationship between the coarse-grained network and the original network by reconstructing subgraphs or contagion configurations from their coarse-grained representations.
The operator is defined for both super-nodes and contagion configurations as follows:
$(\mathrm{i})$ Reconstruction of super-nodes.
For a super-node $v_i(k)$, the operator $\mathcal{H}\{v_i(k)\}$ retrieves the induced subgraph in original networks that was aggregated into the super-node $v_i(k)$ during the coarse-graining process.
This reconstruction follows an iterative inverse coarse-graining process:
\begin{equation}\label{eq:eqs5}
    v_i(k) \xrightarrow{G_{\Delta_{k}}} \cdots \xrightarrow{G_{\Delta_{n}}} \cdots \xrightarrow{G_{\Delta_{\mathcal{K}}}} \mathcal{H}\{v_i(k)\}.
\end{equation}
where $G_{\Delta_{k}}$ represents the $k$-clique CGN, and $\mathcal{H}\{v_i(k)\}$ corresponds to the subgraph in the original network.
This operator captures the hierarchical structure of the coarse-graining process, linking each super-node back to its original corresponding subgraph.
$(\mathrm{ii})$ Mapping contagion configurations.
For a contagion configuration $\mathcal{M}$,
the operator $\mathcal{H}\{\mathcal{M}\}$ reconstructs the corresponding configuration in the original network.
If the super-node $v_i(k)$ is infected in $\mathcal{M}$, all nodes in the subgraph $\mathcal{H}\{v_i(k)\}$ are assumed to be infected.
Mathematically, this is expressed as:
\begin{equation}
\begin{aligned}
    \mathcal{H}\{\mathcal{M}\} =& \{\mathcal{H}\{v_i(k)\}~|~\forall v_i(k) \in V(k), s(v_i(k)|G_{R=k})= 1 \\ 
    &, s(\mathcal{H}\{v_i(k)\}|G) = v_i^w(k)\}.
\end{aligned}
\end{equation}
where $v_i^w(k)$ is the weight of super-node $v_i(k)$, representing the number of nodes in $\mathcal{H}\{v_i(k)\}$, and $s$ denotes the final outbreak size.

\textbf{Conditions for maintaining contagion dynamics}.
Using formal definitions, we derive two conditions that form the theoretical foundation of the ISCG framework, ensuring that key dynamic properties---such as the final outbreak size and the probability of contagion configurations---are preserved between the original network and its coarse-grained representation.
These conditions are further validated through rigorous proofs in Propositions 1 and 2.

For a contagion process originating from node $v_i$ in the original network $G$, the average number of infected nodes can be expressed as:
\begin{equation}\label{eqs:eq6}
   \langle \rho_i(G) \rangle = \sum_{\mathcal{M'}\in \Omega_i^{G}} P(\mathcal{M'}|G) s(\mathcal{M'}|G), 
\end{equation}
where $\Omega_i^{G}$ is the set of all possible contagion configurations starting from node $v_i$ in $G$, $P(\mathcal{M'}|G)$ represents the probability of configuration $\mathcal{M'}$, and $s(\mathcal{M'}|G)$ denotes the number of nodes infected in configuration $\mathcal{M'}$.
This equation provides the baseline for comparing the contagion dynamics between the original network and its CGN, ensuring that the average infection size remains consistent across both representations.

When the transmission probability $\beta$ satisfies $\beta \geq \hat{\beta}_k$, the contagion process within each $k$-clique becomes complete, as shown in Proposition 1. 
This ensures that the contagion configurations in the coarse-grained network, denoted $G_{R=k}$, map directly to those in the original network $G$:
\begin{equation}
\mathcal{H}\{\Omega_i^{G_{R=k}}\} = \Omega_i^{G}, 
\end{equation}
where $\mathcal{H}\{*\}$ is the inverse coarse-graining operator.
Under this condition, the average number of infected nodes in the original network $G$ can be expressed as:
\begin{equation}\label{eqs:eq11}
 \langle \rho_i(G) \rangle =\sum_{\mathcal{M'}\in \mathcal{H}\{\Omega_i^{G_{R=k}}\}} P(\mathcal{M'}|G) s(\mathcal{M'}|G).
\end{equation}
Similarly, the average number of infected nodes in the $k$-clique CGN is:
\begin{equation}\label{eqs:eq12}
\langle \rho_i(G_{R=k}) \rangle =\sum_{\mathcal{M}\in \Omega_i^{G_{R=k}}} P(\mathcal{M}|G_{R=k}) s(\mathcal{M}|G_{R=k}). 
\end{equation}
For equivalence between the original network and the CGN, two conditions must hold:
\begin{eqnarray}\label{eqs:eq13}
	\left \{
	\begin{aligned}
            s(\mathcal{M}|G_{R=k}) & =  s(\mathcal{M'}|G) \\ 
            P(\mathcal{M}|G_{R=k}) & =  P(\mathcal{M'}|G) 
	\end{aligned}
	\right.
\end{eqnarray}
where $\mathcal{M'}= \mathcal{H}\{\mathcal{M}\}$.
When two conditions are satisfied (as guaranteed by Eqs.~\ref{eq:eq1} and \ref{eq:eq2}), the equivalence of average infected nodes follows:
\begin{equation}
\langle \rho_i(G) \rangle = \langle \rho_i(G_{R=k}) \rangle.
\end{equation}
This result demonstrates that the ISCG framework accurately preserves the contagion dynamics of the original network in its coarse-grained representation when $\beta \geq \hat{\beta}_k$.
For $\beta < \hat{\beta}_k$, partial transmission configurations within cliques introduce deviations, as such configurations are excluded in the CGN representation.
An example illustrating the accuracy and potential sources of error for the ISCG is provided in the \SIM{}.
This example offers additional insights into the factors influencing precision during the coarse-graining process.

\textbf{Validation of contagion dynamics consistency}.
The two conditions described in Eqs.~\ref{eq:eq1} and \ref{eq:eq2} ensure the equivalence of two critical properties: the final outbreak size and the probability of contagion configurations.
To validate this consistency, we systematically examine each component of Eq.~\ref{eqs:eq13}:
$(\mathrm{i})$ the final outbreak size is identical between the CGN and the original network, and
$(\mathrm{ii})$ the contagion configuration probabilities are preserved across both representations.

\textit{Equivalence of final outbreak size}.
The first term in Eq.~\ref{eqs:eq13} ensures that the final outbreak size in CGN matches that in the original network under $\beta \geq \hat{\beta}_k$.

The total number of infected nodes for a contagion configuration $\mathcal{M}$ in the CGN is given by:
\begin{equation}\label{eqs:eq14}
    s(\mathcal{M}|G_{R=k}) = \sum_{v_i(k) \in V(k)}s(v_i(k)|G_{R=k})v_i^w(k),
\end{equation}
where $V(k)$ is the set of super-nodes in the $k$-clique CGN.
$s(v_i(k)|G_{R=k})$ is an indicator function: 1 if the super-node $v_i(k)$ is infected in $\mathcal{M}$, and 0 otherwise.
$v_i^w(k)$ is the weight of the super-node $v_i(k)$, representing the number of nodes aggregated from the original network.
This formulation computes the total infected nodes by summing the weights of all infected super-nodes.

For the corresponding contagion configuration $\mathcal{M'}$ in the original network $G$, the final outbreak size is:
\begin{equation}\label{eqs:eq15}
    s(\mathcal{M'}|G) = s(\mathcal{H}\{\mathcal{M}\}|G) = \sum_{v_i(k) \in V(k)}s(\mathcal{H}\{v_i(k)\}|G) 
\end{equation}
where $\mathcal{H}\{v_i(k)\}$ is the subgraph corresponding to the super-node $v_i(k)$ reconstructed via the inverse coarse-graining operator $\mathcal{H}$.
$s(\mathcal{H}\{v_i(k)\}|G)$ is the number of infected nodes within the subgraph $\mathcal{H}\{v_i(k)\}$.

From Proposition 1, when $\beta \geq \hat{\beta_k}$, contagion fully spreads within each $k$-clique, ensuring:
$s(\mathcal{H}\{v_i(k)\}|G) = v_i^w(k)$,
which means all nodes in the subgraph corresponding to the super-node $v_i(k)$ are infected.
Substituting this result into $s(\mathcal{M'}|G)$, we obtain:
\begin{equation}
    s(\mathcal{M'}|G) = \sum_{v_i(k) \in V(k)}s(v_i(k)|G_{R=k})v_i^w(k),
\end{equation}
This matches the final outbreak size in the CGN~(Eq.~\ref{eqs:eq14}).
Therefore, under the condition $\beta \geq \hat{\beta}_k$:
\begin{equation}
s(\mathcal{M}|G_{R=k}) =  s(\mathcal{M'}|G).
\end{equation}
This result demonstrates that the ISCG framework accurately preserves the final outbreak size between the coarse-grained representation and the original network, provided that the transmission probability $\beta$ exceeds the threshold $\hat{\beta}_k$.

\textit{Equivalence of contagion configuration probabilities}.
The second condition in Eq.~\ref{eqs:eq13} ensures that the probability of a contagion configuration $\mathcal{M}$ in the CGN matches that of the corresponding configuration $\mathcal{M'}$ in the original network.

The probability of forming a contagion configuration $\mathcal{M}$ in the CGN is given by:
\begin{widetext}
\begin{equation}\label{eqs:eq16}
\begin{aligned}
  P(\mathcal{M}|G_{R=k}) = & \sum_{\mathcal{P}\in \Psi(\mathcal{M})}\prod \limits_{(v_i(k)\rightarrow v_j(k)) \in \mathcal{P}}P(v_i(k)\rightarrow v_j(k)|G_{R=k},\beta_{w_{ij}}(k)) \\
& \prod \limits_{(v_m(k)\nrightarrow v_n(k)) \in \mathcal{P}}  1 - P(v_m(k)\rightarrow v_n(k)|G_{R=k},\beta_{w_{mn}}(k)),
\end{aligned}
\end{equation}
\end{widetext}
where $\Psi(\mathcal{M})$ is the set of all possible realizations that lead to configuration $\mathcal{M}$, and 
$\mathcal{P}$ represents a sequence of successful ($v_i(k) \rightarrow v_j(k)$) and unsuccessful ($v_m(k) \nrightarrow v_n(k)$) transmissions between super-nodes.

For the corresponding configuration $\mathcal{M'}$ in the original network, the probability is:
\begin{widetext}
\begin{equation}\label{eqs:eq17}
\begin{aligned}
  P(\mathcal{M'}|G) = & \sum_{\mathcal{P'}\in \Psi(\mathcal{H}\{\mathcal{M}\})}\prod \limits_{(\mathcal{H}\{v_i(k)\}\rightarrow \mathcal{H}\{v_j(k)\}) \in \mathcal{P'}}P(\mathcal{H}\{v_i(k)\}\rightarrow \mathcal{H}\{v_j(k)\}|G,\beta) \\
& \prod \limits_{(\mathcal{H}\{v_m(k)\} \nrightarrow \mathcal{H}\{v_n(k)\})\in \mathcal{P'}}  1 - P(\mathcal{H}\{v_m(k)\}\rightarrow \mathcal{H}\{v_n(k)\}|G,\beta),
\end{aligned} 
\end{equation}
\end{widetext}
where $\Psi(\mathcal{H}\{\mathcal{M}\})$ represents all possible realizations that lead to configuration $\mathcal{M}$ in the original network, and $\mathcal{P'}$
denotes sequences of successful and unsuccessful transmissions between subgraphs $\mathcal{H}\{v_i(k)\}$ and $\mathcal{H}\{v_j(k)\}$, which map directly from the CGN configuration $\mathcal{M}$ via inverse coarse-graining.
 
According to Proposition 2, the transmission probabilities between super-nodes in CGN and their corresponding subgraphs in the original network
are equivalent when $\beta \geq \hat{\beta}_k$:
\begin{equation}\label{eq:eqs9}
\begin{aligned}
& P(\mathcal{H}\{v_i(k)\}\rightarrow \mathcal{H}\{v_j(k)\} | G,\beta) \\
& = P(v_i(k) \rightarrow v_j(k)|G_{R=k},\beta_{w_{ij}(k)}).
\end{aligned}
\end{equation}
Substituting this equivalence into the above expressions, we find:
\begin{equation}
P(\mathcal{M'}|G) = P(\mathcal{M}|G_{R=k}).
\end{equation}
This result confirms that ISCG accurately preserves the probability distribution of contagion configurations in the coarse-grained network, ensuring dynamic equivalence with the original network under $\beta \geq \hat{\beta}_k$.
By validating both components of Eq.~\ref{eqs:eq13}, we confirm that the CGN faithfully captures the contagion dynamics of the original network while significantly enhancing computational efficiency.

\subsection*{Proposition 1} 
For any node $v_i(k) \in V(k)$, if $\beta \geq \hat{\beta}_k$ and $s(v_i(k)|G_{R=k}) = 1$, then $s(\mathcal{H}\{v_i(k)\}|G) = v_i^w(k)$, where $v_i^w(k)$ denotes the size of the subgraph in the original network corresponding to $v_i(k)$.

\textbf{Proof}. 
The proposition applies to all nodes in the CGN, regardless of whether they correspond to a single node or aggregate multiple nodes from the original network.
We consider two cases:

Case 1: $v_i^w(k)=1$~(single-node subgraph).
If $v_i(k)$ corresponds to a single node in the original network, its weight is $v_i^w(k)=1$ by definition.
In this case, the subgraph $\mathcal{H}\{v_i(k)\}$ consists of a single node.
Since $s(v_i(k)|G_{R=k}) = 1$, it directly follows that $s(\mathcal{H}\{v_i(k)\}|G) = 1  = v_i^w(k)$.
The relation holds trivially.

Case 2: $v_i^w(k)>1$~(multi-node subgraph).
If $v_i(k)$ aggregates multiple nodes,
it is iteratively constructed by applying the coarse-graining operator $\mathcal{F}$ on its corresponding subgraph $\mathcal{H}\{v_i(k)\}$:
\begin{equation}
\mathcal{H}\{v_i(k)\}\xrightarrow{G_{\Delta_{\mathcal{K}}}} \cdots\xrightarrow{G_{\Delta_{n}}}\cdots\xrightarrow{G_{\Delta_{k}}} v_i(k),
\end{equation}
where $G_{\Delta_{n}}$ denotes a clique of size 
$n$, ranging from $k$ to $\mathcal{K}$ (the size of the largest clique in the network).
By reversing this process, $\mathcal{H}\{v_i(k)\}$ can be reconstructed as a union of cliques with sizes $n \in [k,\mathcal{K}]$: 
\begin{equation}\label{eq:eqs7}
    s(\mathcal{H}\{v_i(k)\}|G)  = \sum\limits_{G_{\Delta}\in\Phi} s(G_{\Delta}|\mathcal{H}\{v_i(k)\}),
\end{equation}
where $\Phi$ is the set of all cliques forming $\mathcal{H}\{v_i(k)\}$.
For $\beta \geq \hat{\beta}_k$, by the definition of $\hat{\beta}_k$, the probability of a single seed infecting all $k-1$ other nodes in a $k$-clique equals 1:
\begin{equation}
P(1\rightarrow (k-1)|G_{\Delta_k},\hat{\beta}_k)=1.
\end{equation}
For larger cliques ($n>k$), since $\hat{\beta}_k > \hat{\beta}_n$~(as shown in Fig.~\ref{fig:fig2}), we similarly have: 
\begin{equation}
P(1\rightarrow (n-1)|G_{\Delta_n},\hat{\beta}_k)=1.
\end{equation}
Thus, all nodes in cliques of sizes $n \in [k,\mathcal{K}]$ are fully infected:
\begin{equation}\label{eq:eqs8_revised}
s(G_{\Delta_n}|\mathcal{H}\{v_i(k)\}) = n.
\end{equation}
If $v_i(k)$ is infected in $G_{R=k}$, at least one node $z \in \mathcal{H}\{v_i(k)\}$ is infected from outside $\mathcal{H}\{v_i(k)\}$.
Due to the self-similarity of $\mathcal{H}\{v_i(k)\}$ and the complete infection of its constituent cliques~(Eq.~\ref{eq:eqs8_revised}), all nodes in $\mathcal{H}\{v_i(k)\}$ are subsequently infected.
Therefore: 
\begin{equation}
s(\mathcal{H}\{v_i(k)\}|G)=v_i^w(k).
\end{equation}

In both cases~($v_i^w(k)=1$ and $v_i^w(k)>1$), the relationship $s(\mathcal{H}\{v_i(k)\}|G) = v_i^w(k)$ holds under the conditions $\beta \geq \hat{\beta}_k$ and $s(v_i(k)|G_{R=k}) = 1$, thereby completing the proof.

\subsection*{Proposition 2}
For any edge $e_{ij}(k) \in E(k)$, if $\beta \geq \hat{\beta}_k$ and $\beta_{w_{ij}(k)} = 1-(1-\beta)^{e^w_{ij}(k)}$, then the transmission probabilities between the induced subgraphs in the original network and the super-nodes in the coarse-grained network are equivalent:
\begin{equation}\label{eq:eqs9b}
\begin{aligned}
& P(\mathcal{H}\{v_i(k)\}\rightarrow \mathcal{H}\{v_j(k)\} |G,\beta) \\
& = P(v_i(k) \rightarrow v_j(k)|G_{R=k},\beta_{w_{ij}(k)}),
\end{aligned}
\end{equation}
where $\mathcal{H}\{v_i(k)\}$ and $\mathcal{H}\{v_j(k)\}$ are the induced subgraphs corresponding to the super-nodes $v_i(k)$ and $v_j(k)$,  respectively, in the original network.

\textbf{Proof}.
This proposition applies to all edges in the $k$-clique CGN, where $v_i(k)$ and $v_j(k)$ represent aggregated subgraphs of the original network.
We consider two cases based on the weight $e_{ij}^w(k)$ of the edge $e_{ij}(k)$.
 
Case 1: $e_{ij}^w(k)=1$ (single-edge case).
If $e_{ij}(k)$ corresponds to a single edge in the original network, the weight $e_{ij}^w(k)=1$ and the transmission probability remain $\beta_{w_{ij}(k)}=\beta$.
By definition, the subgraphs reduce to individual nodes:
$\mathcal{H}\{v_i(k)\}=v_i(k)$, and $\mathcal{H}\{v_j(k)\}=v_j(k)$.
Substituting into the proposition equation, we have:
\begin{equation}
 P(v_i(k)\rightarrow v_j(k) | G,\beta) \\
 = P(v_i(k) \rightarrow v_j(k)|G_{R=k},\beta),
\end{equation}
Since the transmission probabilities between the same pair of nodes in $G$ and $G_{R=k}$ are identical, the equivalence holds trivially.

Case 2: $e_{ij}^w(k)>1$ (multi-edge case).
If $e_{ij}(k)$ aggregates multiple edges in the original network, $e_{ij}^w(k)>1$.
Here, the subgraphs $\mathcal{H}\{v_i(k)\}$ and $\mathcal{H}\{v_j(k)\}$ correspond to the sets of nodes aggregated into the super-nodes $v_i(k)$ and $v_j(k)$, respectively.
From Proposition 1, if $\beta \geq \hat{\beta}_k$, all nodes within $\mathcal{H}\{v_i(k)\}$ and $\mathcal{H}\{v_j(k)\}$ are fully infected when the super-nodes $v_i(k)$ and $v_j(k)$ are infected.
Given that there are $e^w_{ij}(k)$ independent transmission paths between the two subgraphs, the overall transmission probability is:
\begin{equation}\label{eq:eqs10}
    P(\mathcal{H}\{v_i(k)\} \rightarrow \mathcal{H}\{v_j(k)\}|G,\beta) =  1-(1-\beta)^{e^w_{ij}(k)}. 
\end{equation}
In the coarse-grained network, the effective transmission probability along the weighted link $e_{ij}(k)$ is defined as:
\begin{equation}
P(v_i(k) \rightarrow v_j(k)|G_{R=k},\beta_{w_{ij}(k)}) = \beta_{w_{ij}(k)}
\end{equation}
Substituting $\beta_{w_{ij}(k)} = 1-(1-\beta)^{e^w_{ij}(k)}$, we find:
\begin{equation}
\begin{aligned}
    P(\mathcal{H}\{v_i(k)\} \rightarrow \mathcal{H}\{v_j(k)\}|G,\beta) = \\
    P(v_i(k) \rightarrow v_j(k)|G_{R=k},\beta_{w_{ij}(k)}). 
\end{aligned}
\end{equation}
Thus, the transmission probabilities are equivalent in the original network and the $k$-clique CGN.
This completes the proof. 

For further clarification, see the illustrative example provided in the Transmission Probability in Weighted CGNs section of \SIM{}.

\subsection*{Numerical simulations}
The methods and numerical simulations used to generate the figures in this manuscript, along with additional supporting results, are described in detail. 
All results are obtained as averages of $1000$
numerical simulations, each initiated with a randomly selected seed node to trigger the contagion process.

\section*{Data availability}
All data supporting this study are publicly available on Mendeley Data~(\url{https://data.mendeley.com/datasets/dd2m5x9tn9/1}) and are detailed in the \SI{}.

\section*{Code availability}
The codes for implementing the iterative structural coarse-graining method and visualization are available on GitHub at \url{https://github.com/LeyangXue/NetworkReduction.git}. 

\section*{Acknowledgements}
This work is supported by the National Natural Science Foundation~(Grant No. 72274020) and the
Fundamental Research Funds for the Central Universities~(Grant No. 2233200016).
L.X. acknowledges the support of the China Scholarship Council Program and the Israeli Sandwich Scholarship.

\section*{Author contributions}
L.X. conducted the numerical simulations, theoretical analysis, method design, and manuscript preparation,
Z.D. contributed through discussions and feedback on the manuscript,
A.Z. supervised the study, conceptualized the framework, contributed to the refinement of the results, and improved the manuscript.

\section*{Competing interests}
The authors declare no competing financial interests.


\clearpage
\appendix
\setcounter{page}{1}
\renewcommand{\figurename}{Supplementary Figure}
\renewcommand{\tablename}{Supplementary Table}
\setcounter{figure}{0}   
\setcounter{table}{0} 
\numberwithin{equation}{section}

\begin{widetext}
\section*{Supplemental Material: Iterative Structural Coarse Graining for Contagion Dynamics in Complex Networks}

Leyang Xue$^{1,2}$, Zengru Di$^{1,2}$, An Zeng$^{*2}$, 
\\\\
1. International Academic Center of Complex Systems, Beijing Normal University, Zhuhai, 519087, China \\
2. School of Systems Science, Beijing Normal University, Beijing, 100875, China. 
\end{widetext}

\twocolumngrid
\section{Networks}
We apply the coarse-grained method to the selection of real-world networks obtained from the following sources:
(1) Network Repository - NetRes~\cite{Ryan2015};
(2) Github;
(3) Stanford Network Analysis Project - SNAP~\cite{Leskovec2014};
(4) Index of Complex Network - ICN;
(5) Network Science by Albert-Laszlo Barabasi - Barabasi~\cite{Barabasi2016};
These networks capture connectivity patterns across various domains, including scientific collaboration, communication, and social media. 
In each network, nodes represent entities (e.g., scientists, email accounts), with links defining relationships between them. 
To ensure transmission reaches all nodes, we retain only the largest connected component in each network as the base structure.
Detailed characteristics of these networks are presented in Supplementary Table~\ref{tab:stab1}.

\begin{table*}[!htbp]
    \caption{\textbf{Properties of real-world networks}.
     We report the following characteristics for each network:
     network name, number of nodes~($N$), number of edges~($E$), average degree~($\langle k \rangle$), average clustering coefficient~($\langle  c \rangle$), degree correlation coefficient~($r_a$), degree exponent~($\gamma$), estimated by the maximum likelihood method, k-clique exponent in the original network~($\alpha_k$), k-clique exponent generated during coarse-graining process~$\alpha_{\Delta k}$, and the networks source.}
    \label{tab:stab1}
    \begin{tabular}{rrrrrrrrrr}\hline
    \textbf{Network} & \textbf{$N$} & \textbf{$E$} & \textbf{$\langle k \rangle$} & \textbf{$\langle  c \rangle$} & \textbf{$r_a$} & \textbf{$\gamma$} & \textbf{$\alpha_k$} &\textbf{$\alpha_{\Delta k}$} & \textbf{Source}\\ \hline
    AstroPh & 17,903 & 196,972 & 22.00 & 0.63 & 0.20 & 4.50 & -7.36 & -3.00 & NetRes\\
    CondMat & 21,363 & 91,286 & 8.55 & 0.64 & 0.13 & 3.35 & -6.50 & -4.03 & NetRes\\
    DBLP & 12,495 & 49,563 & 7.93 & 0.12 & -0.05 & 3.35 & -5.39 & -4.91 & NetRes\\
    Deezer-europe & 28,281 & 92,752 & 6.56 & 0.14 & 0.10 & 4.86 & -8.13 & -6.10 & Github\\
    Email-Enron & 33,696 & 180,811 & 10.73 & 0.51 & -0.12 & 1.97 & -11.40 & -7.08 & NetRes\\
    GrQc & 4,158 & 13,422 & 6.46 & 0.56 & 0.64 & 2.04 & -3.78 & -5.84 & NetRes\\
    HepPh & 11,204 & 117,619 & 21.00 & 0.62 & 0.63 & 2.08 & -2.32 & -3.85 & NetRes\\
    Musae-DE & 9,498 & 153,138 & 32.25 & 0.20 & -0.12 & 2.55 & -11.73 & -7.28 & SNAP\\
    Musae-facebook & 22,470 & 170,823 & 15.20 & 0.36 & 0.08 & 3.19 & -8.18 & -4.04 & SNAP\\
    Musae-git & 37,700 & 289,003 & 15.33 & 0.17 & -0.08 & 2.54 & -8.72 & -5.89 & SNAP\\
    NetSci2019 & 32,904 & 296,876 & 18.04 & 0.80 & 0.99 & 2.02 & -4.34 & -4.36 & ICN\\
    Phonecalls & 30,420 & 52,841 & 3.47 & 0.15 & 0.17 & 4.71 & -5.67 & -8.12  & Barabasi\\ \hline
    \end{tabular}
\end{table*}

\section{Supplementary Methods}
\label{SI:method}

\subsection*{Recursive calculation of full infection probability in \texorpdfstring{\(k\)}{k}-cliques}

\label{SI:occupy}

\begin{figure*}[!htbp]
	\centering
	\includegraphics[width=\textwidth]{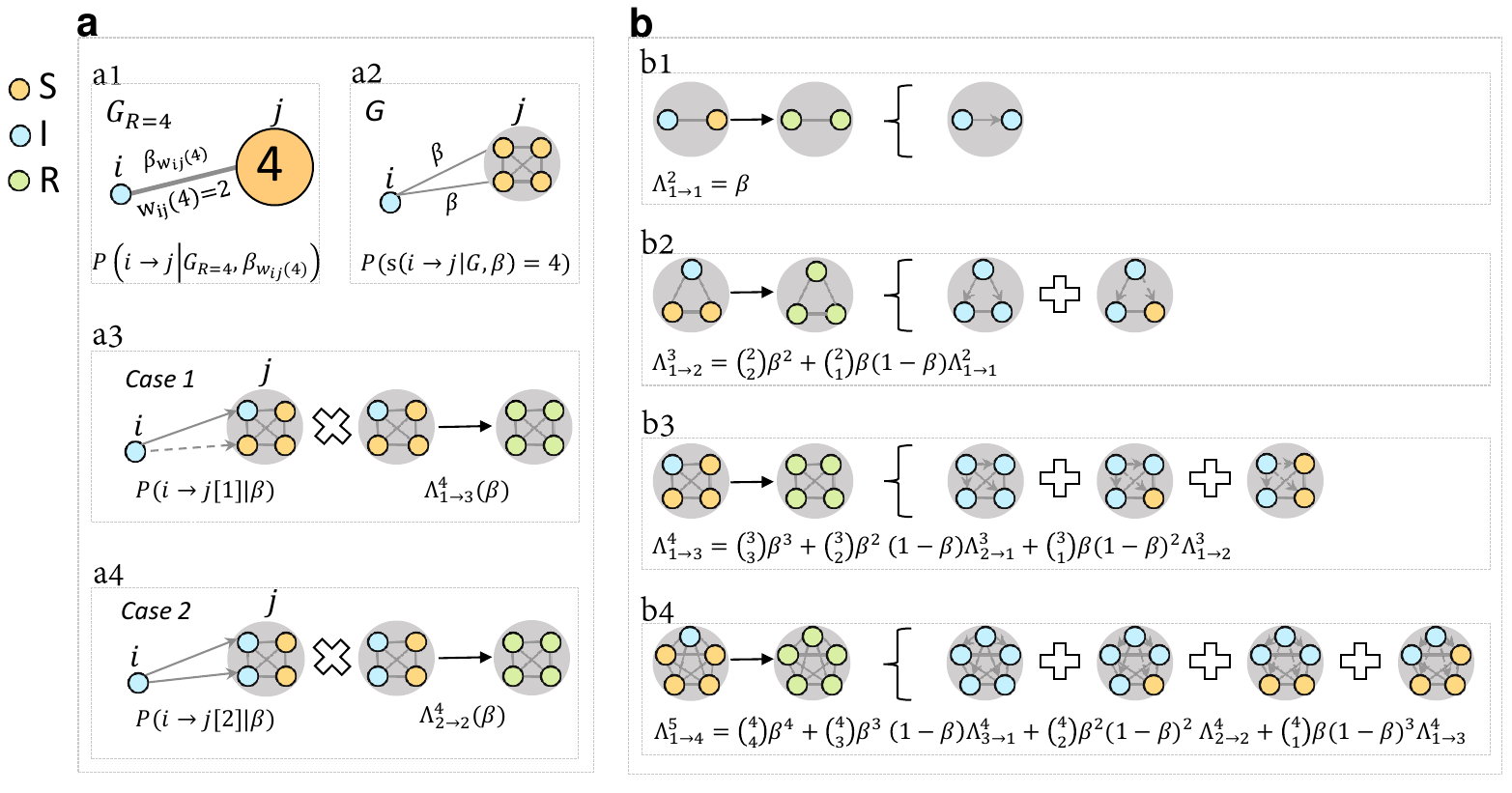}
	\caption{\textbf{Illustration of infection probabilities in coarse-grained networks~(CGNs) and $k$-clique structures}.
    \textbf{a} Probability mapping in $4$-clique CGNs.
    \textbf{a1} Example of a 4-clique CGN showing node $i$ infecting node $j$ through an edge with weight 2.
    \textbf{a2} Corresponding contagion configuration in the original network, 
    highlighting the infection of all nodes within node $j$’s clique. 
    \textbf{a3} Case 1: Node $i$ infects one node in $j$’s clique via one of the two links, and the infected node spreads the contagion to all remaining nodes in the \(4\)-clique.
    \textbf{a4} Case 2: Node $i$ infects two nodes simultaneously, with the infected nodes subsequently spreading the contagion to the rest of the \(4\)-clique.
    \textbf{b} Infection dynamics in $k$-clique structures (for $k =2$ to $5$):
    \textbf{b1-b4} Scenarios illustrating how a single seed node infects all other nodes in a $k$-clique.
    The left side of each brackets shows the initial seed infecting all nodes, while  the right side lists possible configurations enabling this event.
    Solid arrows within the clique denote successful transmissions, and dashed arrows represent failed attempts. }
	\label{fig:figS1}
\end{figure*}

In this section, we present the recursive framework used to calculate the probability that a single initially infected node within a
\( k \)-clique ultimately spreads the infection to all other nodes, denoted as \(\Lambda^k_{1 \rightarrow (k-1)}(\beta)\), where \(\beta\) is the infection probability between any two connected nodes.
This approach allows for precise estimation of infection probabilities by systematically analyzing each possible contagion pathway within the clique structure.

\subsubsection{General recursive formula}
To compute \(\Lambda^k_{1 \rightarrow (k-1)}(\beta)\), we divide the infection process into sequential infection stages. 
Starting with a single infected node, we recursively calculate the probability of the infection spreading to all \( k-1 \) remaining nodes. 
The recursive formula is defined as follows:
\begin{equation}
\Lambda^k_{1 \rightarrow (k-1)}(\beta) = \sum_{i=1}^{k-1} \binom{k-1}{i} \beta^i (1 - \beta)^{k-1-i} \Lambda^{k-1}_{i \rightarrow (k-1-i)}(\beta)
\end{equation}
where:
\begin{itemize}
    \item \(i\) is the number of nodes initially infected by the seed in the next step,
    \item \(\binom{k-1}{i}\) denotes the number of ways to choose \(i\) nodes from the \(k-1\) nodes,
    \item \(\beta^i (1 - \beta)^{k-1-i}\) is the probability that exactly \(i\) nodes are infected while \(k-1-i\) remain uninfected,
    \item \(\Lambda^{k-1}_{i \rightarrow (k-1-i)}(\beta)\) represents the probability that the \(i\) infected nodes can propagate the infection to the remaining \(k-1-i\) nodes in a \( (k-1)\)-clique.
\end{itemize}

\subsubsection{Example calculations for small cliques}
To illustrate the recursive framework, we provide specific calculations for cliques of size \(k=2\) through \(k=5\).
For a 2-clique, which consists of two nodes connected by a single edge, the probability of full infection is straightforward:
\begin{equation}
	\begin{aligned}
	\Lambda^2_{1\rightarrow1}(\beta) =\tbinom{1}{1} \beta \Lambda^1_{1\rightarrow 0}(\beta)=\beta.
	\end{aligned}
\end{equation}

In a 3-clique, we consider two possible pathways for full infection, as illustrated in Supplementary Figure~\ref{fig:figS1}~(b2): (i) the initial seed directly infects both remaining nodes or (ii) the seed infects one node, which subsequently infects the third node. 
This yields:
\begin{equation}
	\begin{aligned}
	\Lambda^3_{1\rightarrow 2}(\beta) &= \tbinom{2}{2}\beta^2(1-\beta)^0\Lambda^2_{2\rightarrow0}(\beta)+\tbinom{2}{1}\beta(1-\beta)\Lambda^2_{1\rightarrow1}(\beta)\\ 
    &= -2 \beta^3+3 \beta^2.
	\end{aligned}
\end{equation}

For a 4-clique, we enumerate the cases where one to three nodes are infected in the next step, using the recursive formula to account for further infections~(see Supplementary Figure\ref{fig:figS1}~(b3)). 
This yields:
\begin{equation}\label{eq:eq4}
	\begin{aligned}
	\Lambda^4_{1\rightarrow3}(\beta)= &\tbinom{3}{3}\beta^3(1-\beta)^0\Lambda^3_{3\rightarrow0}(\beta)\\
     + & \tbinom{3}{2}\beta^2 (1-\beta) \Lambda^3_{2\rightarrow1}(\beta) + \tbinom{3}{1}\beta(1-\beta)^2 \Lambda^3_{1\rightarrow2}(\beta)\\
    = & 24\beta^5 - 33\beta^4 + 16\beta^3 - 6\beta^2, 	
	\end{aligned}
\end{equation}
where $\Lambda^3_{2\rightarrow1}(\beta)$ denotes the probability that two infected nodes succeed in infecting an additional susceptible node within a \(3\)-clique, given by $1-\tbinom{1}{0}\Lambda^2_{2\rightarrow0}(\beta)(1-\beta)^{2\times1}(1-\beta)^{0\times1}$, accounting for the complementary probability of no further infections.

In the 5-clique, we follow a similar enumeration process, as shown in Supplementary Figure\ref{fig:figS1}~(b4):
\begin{equation}
	\begin{aligned}
		\Lambda^5_{1\rightarrow4}(\beta) = & \tbinom{4}{4}\beta^4(1-\beta)^0\Lambda^4_{4\rightarrow0}(\beta)+\tbinom{4}{3}\beta^3(1-\beta)\Lambda^4_{3\rightarrow1}(\beta)\\
		+&\tbinom{4}{2}\beta^2(1-\beta)^2\Lambda^4_{2\rightarrow2}(\beta)+\tbinom{4}{1}\beta(1-\beta)^3\Lambda^4_{1\rightarrow3}(\beta)\\
		=& 24\beta^{10}-180\beta^9+570\beta^8-980\beta^7+970\beta^6 \\
        -& 528\beta^5 + 125\beta^4.
	\end{aligned}
\end{equation}
In this framework, $\Lambda^4_{3\rightarrow1}(\beta)$ represents the probability that three nodes will infect single susceptible node in a \(4\)-clique, calculated as $1-\Lambda^3_{3\rightarrow0}(\beta)(1-\beta)^{3}$. 
$\Lambda^4_{2\rightarrow2}(\beta)$ represents the probability that two infected nodes will infect two susceptible nodes in a \(4\)-clique, considering all complementary (non-infection) cases, defined by $\Lambda^4_{2\rightarrow2}(\beta) =1-[\tbinom{2}{0}\Lambda^2_{2\rightarrow0}(\beta)(1-\beta)^{2\times2}+\tbinom{2}{1}\Lambda^3_{2\rightarrow1}(\beta)(1-\beta)^{2\times1}(1-\beta)]$.
This iterative framework allows for a clear, step-by-step computation of infection probabilities within \(k\)-clique networks.


\subsubsection{Extending to larger cliques}
Using this recursive formula, we extend the approach to larger cliques. 
Each \(k\)-clique builds upon the infection probabilities of smaller cliques by recursively applying infection pathways and summing the probabilities for each possible configuration.
This recursive framework captures all potential pathways of infection within a 
\( k \)-clique, providing a scalable approach to calculate complete infection probabilities across densely connected networks. 
This approach offers insight into the intricate dynamics of contagion in tightly interconnected structures and is well-suited for studying contagion in dense network environments.

The probability of complete infection for a 
\(6\)-clique is given by: 
\begin{equation}
	\begin{aligned}
    \Lambda^6_{1\rightarrow5}(\beta) = & \sum_{i=1}^5\tbinom{5}{i}\beta^i(1-\beta)^{5-i}\Lambda^5_{i\rightarrow 5-i}(\beta)\\
	= & 120\beta^{15} - 1440\beta^{14} + 7830\beta^{13} - 25440\beta^{12} + \\
    & 54780\beta^{11} - 81840\beta^{10} + 86110\beta^{9}- 63195\beta^{8} + \\
    & 31080\beta^{7} - 9300\beta^{6} + 1296\beta^{5},
	\end{aligned}
\end{equation}
where each term $\Lambda^5_{i\rightarrow(5-i)}(\beta)$ represents the probability of subsequent infection for each configuration in the 5-clique.

Similarly, for a \(7\)-clique, the probability expression is:
\begin{equation}
\begin{aligned}
\Lambda^7_{1\rightarrow6}(\beta) =& \sum_{i=1}^6\tbinom{6}{i}\beta^i(1-\beta)^{6-i}\Lambda^6_{i\rightarrow 6-i}(\beta) \\
		 = & - 720\beta^{21} + 12600\beta^{20} - 103320\beta^{19} + 526890\beta^{18}\\ 
            & - 1869840\beta^{17} + 4894680\beta^{16}- 9770810\beta^{15} \\
            & + 15159120\beta^{14} - 18448710\beta^{13} + 17633945\beta^{12} \\
            & - 13150032\beta^{11}+ 7527471\beta^{10} - 3209430\beta^{9} \\
            & + 965160\beta^{8} - 183810\beta^{7} + 16807\beta^{6},
\end{aligned}
\end{equation}
where each term $\Lambda^6_{i\rightarrow 6-i}(\beta)$ is computed recursively based on previous results for the \(6\)-clique structure.
For conciseness, we can reference the earlier formulas for $\Lambda^6_{i\rightarrow 6-i}(\beta)$ and $\Lambda^5_{i\rightarrow j}(\beta)$ as needed. 
For clarity, subscripts are used to specify particular values of $i$ and $j$.
For instance, we denote $\Lambda^6_{5\rightarrow1}(\beta)$ represents the probability that five infected nodes successfully infect the one remaining susceptible node in a 6-clique, calculated as: $\Lambda^6_{5\rightarrow1}(\beta) = 1-\tbinom{1}{0}\Lambda^5_{5\rightarrow0}(\beta)(1-\beta)^{5\times1}(1-\beta)^{0\times1}$.

This provides a recursive framework to compute the full infection probability for \(k\)-clique networks of increasing size. 
For $k=8$, $k=9$, and $k=10$, we leverage this framework to express infection probabilities as functions of $\beta$, capturing complex infection dynamics across larger network structures with precision.
\begin{equation}
	\begin{aligned}
		\Lambda^8_{1\rightarrow7}(\beta) &= \sum_{i=1}^{7} \tbinom{7}{i}\beta^{i}(1-\beta)^{7-i}\Lambda^7_{i\rightarrow7-i}(\beta)\\
		& = -5040\beta^{28} + 120960\beta^{27} - 1386000\beta^{26} \\
        & + 10086720\beta^{25} - 52319190\beta^{24} + 205732800\beta^{23} \\
        & - 636845160\beta^{22} + 1590501640\beta^{21} - 3258291120\beta^{20} \\
		& + 5536123600\beta^{19} - 7856193296\beta^{18} + 9345271992\beta^{17} \\
        & - 9324568001\beta^{16} + 7786027816\beta^{15} - 5410382880\beta^{14} \\
        & + 3098951072\beta^{13} - 1441519296\beta^{12}  + 532354536\beta^{11} \\
        & - 150657080\beta^{10} + 30802240\beta^{9} - 4068456\beta^{8} \\
        & + 262144\beta^{7}
	\end{aligned}
\end{equation}

\begin{equation}
	\begin{aligned}
		\Lambda^9_{1\rightarrow8}(\beta) &= \sum_{i=1}^{8} \tbinom{8}{i}\beta^{i}(1-\beta)^{8-i}\Lambda^8_{i\rightarrow8-i}(\beta)\\
		& = 40320\beta^{36} - 1270080\beta^{35} + 19323360\beta^{34} \\
        & - 189090720\beta^{33} + 1337084280\beta^{32} \\
        & - 7276802400\beta^{31}  + 31699531080\beta^{30} \\
        & - 113494862880\beta^{29} + 340283488860\beta^{28} \\
        & - 866124834680\beta^{27} + 1890561025584\beta^{26} \\
        & - 3565563257088\beta^{25}  + 5841996257286\beta^{24} \\
        & - 8346992077872\beta^{23} + 10424251015920\beta^{22} \\
        & - 11390487268104\beta^{21} + 10886907107538\beta^{20} \\
        & - 9087542239104\beta^{19}  + 6604921452864\beta^{18} \\
        & - 4160649753288\beta^{17} + 2256642859464\beta^{16}\\
		& - 1044267039720\beta^{15} + 407158789500\beta^{14}\\
        & - 131447465856\beta^{13}  + 34273176588\beta^{12} \\
        & - 6952118544\beta^{11}+ 1032288516\beta^{10}\\ 
        & - 100143792\beta^{9} + 4782969\beta^{8}\\
	\end{aligned}
\end{equation}

\begin{widetext}
\begin{equation}
	\begin{aligned}
		\Lambda^{10}_{1\rightarrow9}(\beta) &= \sum_{i=1}^{9} \tbinom{9}{i}\beta^{i}(1-\beta)^{9-i}\Lambda^9_{i\rightarrow9-i}(\beta)\\
		&=362880\beta^{45} - 14515200\beta^{44} + 282592800\beta^{43} - 3567715200\beta^{42} + 32833495800\beta^{41}\\
		& - 234748765440\beta^{40} + 1357020856800\beta^{39} - 6517548349200\beta^{38} + 26521978127400\beta^{37} \\
		& - 92792729053500\beta^{36} + 282287441908080\beta^{35} - 753273866698920\beta^{34}\\ 
		& + 1775448575926410\beta^{33} - 3716558335019880\beta^{32} + 6939551178972720\beta^{31}\\
		& - 11596879696617600\beta^{30} + 17388982649046960\beta^{29} - 23437879996999860\beta^{28}\\ 
		& + 28429756177413360\beta^{27}- 31050312703343640\beta^{26} + 30532209914200806\beta^{25}\\
		& - 27011077082801580\beta^{24} + 21469710851551800\beta^{23} - 15300758477189520\beta^{22}\\
		& + 9748958193896580\beta^{21} - 5532426738592740\beta^{20} + 2782630494934920\beta^{19}\\ 
		& - 1232671556293800\beta^{18} + 477077447178540\beta^{17} - 159642667620135\beta^{16} \\
		& + 45558310696800\beta^{15}  - 10884316965480\beta^{14} + 2121183237600\beta^{13}\\ 
		& - 324496267200\beta^{12} + 36628300800\beta^{11} - 2719892160\beta^{10} + 100000000\beta^{9}\\
	\end{aligned}
\end{equation}
\end{widetext}

Each term $\Lambda^{k}_{i\rightarrow (k-i)}$ reflects the probability that $i$ infected nodes continue to spread the infection to remaining susceptible nodes in a \(k\)-clique
configuration.

\subsubsection{Threshold for complete contagion in \texorpdfstring{$k$}{k}-cliques}
To achieve full infection within cliques, we identify the minimum transmission probability, $\hat{\beta}_k$, at which $\Lambda^k_{1\rightarrow(k-1)}(\beta)=1$.
This threshold ensures that a single infected node can successfully transmit the infection to all other nodes in \(k\)-cliques.
Consequently, $\hat{\beta}_k$ is a critical parameter for enabling complete infection within cliques, thereby supporting the accurate representation of network dynamics in the ISCG framework.

The concept of $\hat{\beta}_k$ parallels the fixation probability in birth-death processes, where the system reaches an absorbing state once a critical threshold is surpassed.
Similarly, $\hat{\beta}_k$ acts as a tipping point, ensuring that the ISCG framework effectively preserves the contagion dynamics during coarse-graining.
By meeting this condition, the framework maintains accuracy and consistency across multiple scales, allowing for reliable analysis of network behavior.

\subsection*{Transmission probability in weighted CGNs}
\label{SI:weight}
In coarse-grained networks~(CGNs), super-edges are assigned weights based on the number of edges aggregated from the original network. 
To ensure that contagion dynamics are accurately preserved in the CGN representation, these weights must be effectively mapped to infection probabilities that reflect the likelihood of transmission between super-nodes. 
This mapping is crucial for maintaining the fidelity of the contagion process at coarser network scales.

\textbf{Mapping edge weights to probabilities}.
The infection probability $\beta_{w_{ij}(k)}$ for an edge with weight $w_{ij}(k)$ in a $k$-clique CGN is defined as:
\begin{equation}
\beta_{w_{ij}}(k) = 1-(1-\beta)^{w_{ij}(k)},
\end{equation}
where $\beta$ is the infection probability on a single edge in the original network. 
This formula accounts for all independent infection paths represented by the aggregated edges, ensuring that the cumulative transmission probability between two super-nodes reflects the contributions of each edge in the original network.

An example is provided in Supplementary Figure~\ref{fig:figS1}, which illustrates the calculation of infection probabilities in a 4-clique CGN. 
In Supplementary Figure~\ref{fig:figS1} a1, the edge between node $i$ and node $j$ in the CGN has a weight of 2, representing two independent transmission pathways in the original network.
The corresponding infection probability $\beta_{w_{ij}(k)}$ is computed using the formula, capturing the cumulative likelihood of transmission.
This mapping is fundamental to the ISCG framework.
By ensuring that $\beta_{w_{ij}(k)}$ accurately reflects the aggregated transmission pathways, the CGN can reliably model contagion processes even at coarse-grained scales. 
This guarantees the preservation of the original network's dynamic behavior while enabling significant reductions in computational complexity and storage requirements.
  
\subsection*{Accuracy analysis of ISCG framework}
\label{SI:analysis}
\begin{figure*}[!htbp]
	\centering
	\includegraphics[width=\textwidth]{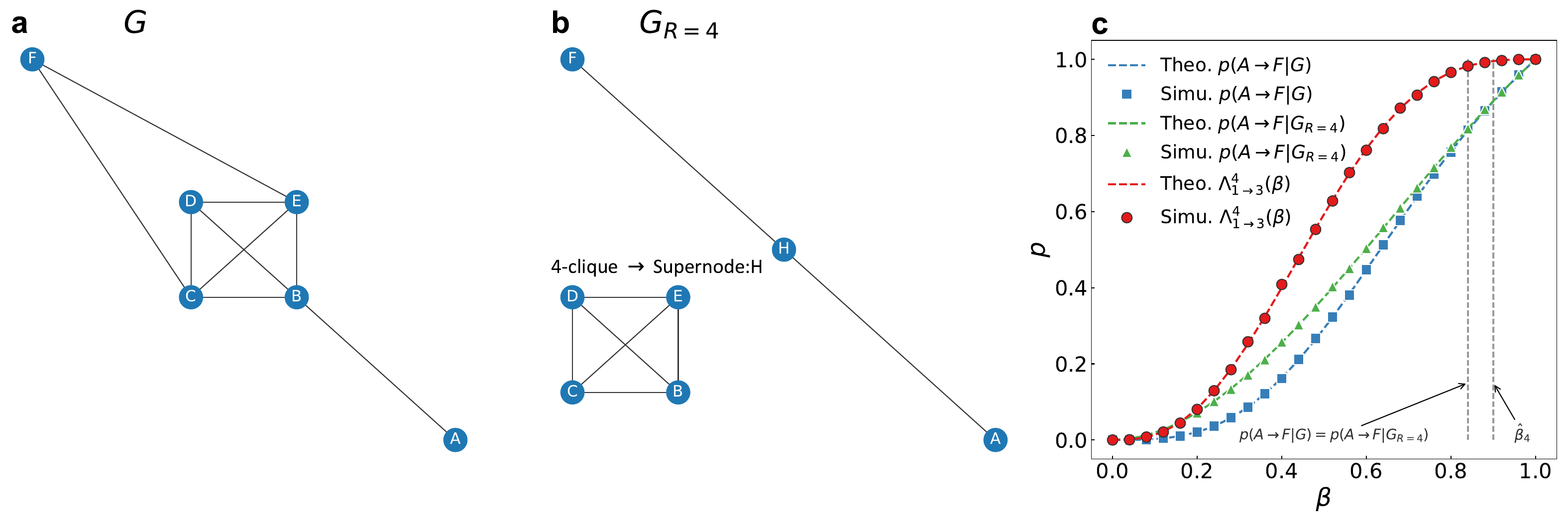}
        \caption{\textbf{Validation of the ISCG framework in preserving SIR dynamics.}
        \textbf{a} A toy network $G$, consisting of two isolated nodes ($A$ and $F$) and a 4-clique substructure~($B$, $C$, $D$, and $E$).
        \textbf{b} The corresponding 4-clique CGN $G_{R=4}$ where the 4-clique is reduced to a super-node~($H$) through the ISCG process.
        \textbf{c} The probability of node $A$ infecting node $F$ in the original network~(G) and the 4-clique CGN~($G_{R=4}$) is plotted as a function of the transmission probability $\beta$. 
        The figure also shows the probability $\Lambda^{4}_{1\rightarrow3}(\beta)$, representing the likelihood of a single seed infecting all other nodes in the 4-clique.
        Simulation results~(markers) are compared with theoretical predictions~(dashed lines), confirming strong agreement.
        Vertical dashed lines indicate key thresholds:  transmission probability $\hat{\beta_4}$ where $\Lambda^{4}_{1\rightarrow3} =1$, and the point where $P(A \rightarrow F|G) = P(A \rightarrow F|G_{R=4})$, demonstrating the accuracy of the ISCG framework in preserving contagion dynamics.
        }
	\label{fig:figS2}
\end{figure*}

The accuracy of the ISCG framework in preserving contagion dynamics lies in its ability to faithfully represent the contagion process in coarse-grained networks (CGNs). 
To demonstrate this, we analyze a simple example involving a toy network $G$ with six nodes: a single 4-clique~($B$, $C$, $D$, $E$) connected to nodes $A$ and $F$~(Supplementary Figure~\ref{fig:figS2} a).
Using this example, we explicitly compute the contagion probability $P(A \rightarrow F|G)$ in the original network and $P(A \rightarrow F|G_{R=4})$ in its 4-clique CGN representation (Supplementary Figure~\ref{fig:figS2} b). 
This analysis highlights the conditions under which the ISCG framework accurately captures the contagion dynamics and identifies potential sources of error.

\textbf{Contagion in the original network}.
The calculation of $P(A \rightarrow F|G)$ in the original network $G$ involves enumerating all possible contagion pathways through the 4-clique structure.
These pathways can be categorized into three distinct types: 
(1) Complete clique infection pathway. 
\begin{itemize}
\item $A$ infects a node in the clique~(i.e., $B$).
\item The contagion spreads all nodes in the 4-clique ($B$, $C$, $D$, $E$).
\item At least one infected node within the clique subsequently infects $F$.
\end{itemize}
For this pathway, the probability depends on three factors: $(\mathrm{i})$ the probability of $A$ infecting $B$, $(\mathrm{ii})$ the probability of the complete infection of the 4-clique~($\Lambda^4_{1\rightarrow3}(\beta)$, Eq.~\ref{eq:eq4}) and $(\mathrm{iii})$ the probability of the clique infecting $F$. 
The resulting expression is:
\begin{equation}
P_{complete}(A \rightarrow F|G) = \beta \Lambda^4_{1\rightarrow3}(\beta)(1-(1-\beta)^2).
\end{equation}
(2) Partial clique infection pathway: single node.
\begin{itemize}
\item $B$ infects one node in the clique~(e.g., $C$ or $E$), but the contagion does not spread to the rest of the clique.
\item The infected node directly infects $F$.
\end{itemize}
The probability for this pathway accounts for the direct transmission through a single node in the clique:
\begin{equation}
P_{single}(A \rightarrow F|G) = 2\beta^3(1-\beta)^4.
\end{equation}
(3) Partial clique infection pathway: multiple nodes.
\begin{itemize}
\item $B$ infects two nodes in the clique~(e.g., $C$ and $D$ or $E$ and $D$). 
\item These infected nodes attempt to infect $F$, but the remaining nodes in the clique remain uninfected.
\end{itemize}
The probability for this pathway considers all configurations where multiple nodes in the clique are infected:
\begin{equation}
P_{multiple}(A \rightarrow F|G) = 2\beta^2\Lambda^3_{1\rightarrow 2}(\beta)(1-\beta)^3
\end{equation}
Summing these pathways, the total contagion probability $P(A \rightarrow F|G)$ in the original network is:
\begin{equation}
P(A \rightarrow F|G) = 4\beta^9 - 19\beta^8 + 32\beta^7 - 19\beta^6 - 3\beta^5 + 4\beta^4 + 2\beta^3
\end{equation}

\textbf{Contagion in the coarse-grained network}.
In the 4-clique $G_{R=4}$,  the internal dynamics of the 4-clique are abstracted into a single super-node $H$.
The contagion process simplifies into two steps: (1) $A$ infects $H$ with probability $\beta$, and (2) $H$ infects $F$ with an effective probability $\beta(1-(1-\beta)^2)$.
The total contagion probability is:
\begin{equation}
P(A \rightarrow F|G_{R=4}) = 2\beta^2-\beta^3
\end{equation}
This abstraction significantly reduces computational complexity while maintaining accuracy under specific conditions.
 
\textbf{Accuracy and source of error}.
The ISCG framework assumes that if the transmission probability $\beta$ exceeds a critical threshold $\hat{\beta}_4$, the contagion process within the $4$-clique will be complete~(i.e., all nodes in the clique will be infected).
Under this condition, the CGN representation accurately preserves the contagion dynamics of the original network. 
Specifically, $P(A \rightarrow F|G)$, the contagion probability in the original network, becomes equal to $P(A \rightarrow F|G_{R=4})$, the probability in the 4-clique CGN. 
To understand this, we analyze the three distinct infection pathways that contribute to
\begin{equation}
P(A \rightarrow F|G) = P_{complete} + P_{single} + P_{multiple}.
\end{equation}
When $\beta \geq \hat{\beta}_4$, the probabilities $P_{single}$ and $P_{multiple}$ become negligible due to their dependence on terms like $(1-\beta)^3$ and $(1-\beta)^4$, which decay exponentially.
Quantitatively, $P_{single} = 2\beta^3(1-\beta)^4 \rightarrow 0 $ and $P_{multiple} = 2\beta^2\Lambda^3_{1\rightarrow 2}(\beta)(1-\beta)^3 \rightarrow 0$, as $(1-\beta) \rightarrow 0$.
Conversely, $P_{complete}$ dominates as $\Lambda^4_{1\rightarrow 3}(\beta) \rightarrow 1$ and $1-(1-\beta)^2 \rightarrow 1$.
Thus, we have $P(A \rightarrow F|G) \rightarrow P_{complete} = P(A \rightarrow F|G_{R=4})$.
This analysis confirms that for $\beta \geq \hat{\beta}_4$, the ISCG framework preserves contagion dynamics, as the CGN accurately reflects the original network's infection process.

This analysis is further confirmed in Supplementary Figure \ref{fig:figS2} c, where the theoretical and simulated probabilities for both $G$ and $G_{R=4}$ converge at $\beta= \hat{\beta}_4$.
For $\beta < \hat{\beta}_4$, the coarse-grained representation tends to overestimate $P(A \rightarrow F)$, as it assumes that contagion within the clique is always complete.
However, in the original network, partial contagion pathways are more probable under lower transmission probability, leading to discrepancies.
These discrepancies arise from three distinct contagion pathways: $P_{complete}$, $P_{single}$ and $P_{multiple}$.
For $\beta<\hat{\beta_4}$, $P_{single}$ and $P_{multiple}$ dominate in the original network, reflecting scenarios where partial contagion within the clique drives the spread to $F$. 
As $\beta$ increases and approaches $\hat{\beta_4}$, the likelihood of complete clique contagion ($P_{complete}$) becomes dominant, and $P_{single}$ and $P_{multiple}$ diminish to near zero, ensuring that $P(A \rightarrow F|G) = P(A \rightarrow F|G_{R=4})$ at $\beta \geq \hat{\beta_4}$.

Interestingly, we identify an infection probability $\beta_{eq}$, at which $P(A \rightarrow F|G)$ and $P(A \rightarrow F|G_{R=4})$ become equal.
Notably, $\beta_{eq}$ is significantly smaller than the theoretical threshold $\hat{\beta}_4$, indicating that the ISCG framework performs better in practice than predicted by theory. 
This observation highlights the robustness of the ISCG framework, which not only achieves accuracy under critical conditions but also demonstrates enhanced flexibility and effectiveness in preserving contagion dynamics, surpassing the expectations set by theoretical thresholds.

\section{Supplementary Results}
\label{SI:result}

\subsection*{Reduction performance on real-world networks} 
\label{SI:ratio}

\begin{figure*}[!htbp]
	\centering
	\includegraphics[width=\textwidth]{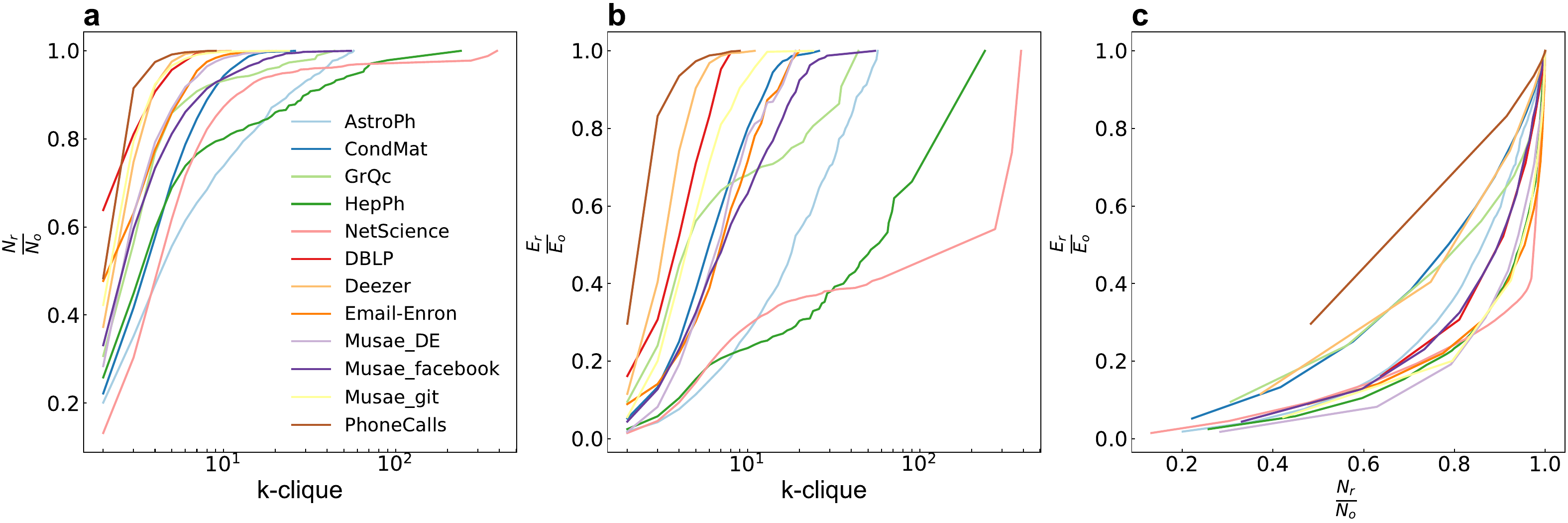}
	\caption{\textbf{Reduced ratio of $k$-clique CGNs for various real-world networks}. 
    \textbf{a, b} Proportion of nodes~($N_r/N_o$) and edges~($E_r/E_o$) in $k$-clique CGNs relative to the original networks as a function of $k$.
    \textbf{c} Relationship between edge reduction ($E_r/E_o$) and node reduction~($N_r/N_o$) as $k$ increases from 2 to $\mathcal{K}$~(i.e. the size of the largest clique in the networks).}
	\label{fig:figS3}
\end{figure*}

In this section, we further evaluate the ISCG framework’s capability in reducing network complexity, and apply it to a diverse set of real-world networks, including AstroPh, DBLP, Deezer, Email-Enron, Musae\_DE, Musae\_facebook, Musae\_git, and PhoneCalls.
The reduction performance for nodes and edges across these datasets is summarized in Supplementary Figures~\ref{fig:figS3} a–c.

\textbf{Node reduction}. 
Supplementary Figure~\ref{fig:figS3} a illustrates the proportion of nodes ($N_r/N_o$) in the $k$-clique CGNs relative to the original networks as a function of $k$. 
Across all networks, the proportion of nodes decreases sharply as $k$ decreases, demonstrating the framework's significant reduction capabilities.
Notably, when reduced to $4$-clique CGNs, the number of nodes in most networks is approximately halved compared to the original network. 
This emphasizes the ISCG framework's effectiveness in simplifying network size, even at relatively fine levels of coarse-graining.

\textbf{Edge reduction}.
The edge reduction is shown in Supplementary Figure~\ref{fig:figS3} b, which presents the proportion of edges ($E_r/E_o$) in the $k$-clique CGNs relative to the original networks. 
The edge reduction is even more pronounced than the node reduction due to the ISCG framework's ability to merge dense subgraphs into super-nodes. 
This results in a substantial simplification of the network's connectivity structure while preserving key topological and dynamic properties.

\textbf{Relationship between node and edge reduction}.
To understand the relationship between node and edge reductions, Supplementary Figure~\ref{fig:figS3}(c) plots $E_r/E_o$ as a function of $N_r/N_o$. 
The results reveal that as the node reduction rate ($N_r/N_o$) decreases from 1, the edge reduction rate ($E_r/E_o$) declines disproportionately, often faster than linear scaling. 
This highlights the framework's effectiveness in targeting and simplifying redundant connections within the network.
The significant reductions in nodes and edges have practical implications for analyzing large-scale networks. 
In many theoretical methods (e.g., the dynamical message-passing algorithm) and numerical simulations, computational complexity scales with the number of edges.
By disproportionately reducing edges, the ISCG framework substantially decreases computational overhead and storage requirements. 
This makes it particularly suitable for studying dynamic processes on large-scale networks, where efficient coarse-graining is essential for scalability and feasibility.
Overall, these results underscore the versatility and efficiency of the ISCG framework in handling a wide variety of network types while maintaining critical dynamic behaviors.

\subsection*{Approximate reduction} 
\begin{figure*}[!htbp]
	\centering
	\includegraphics[width=\textwidth]{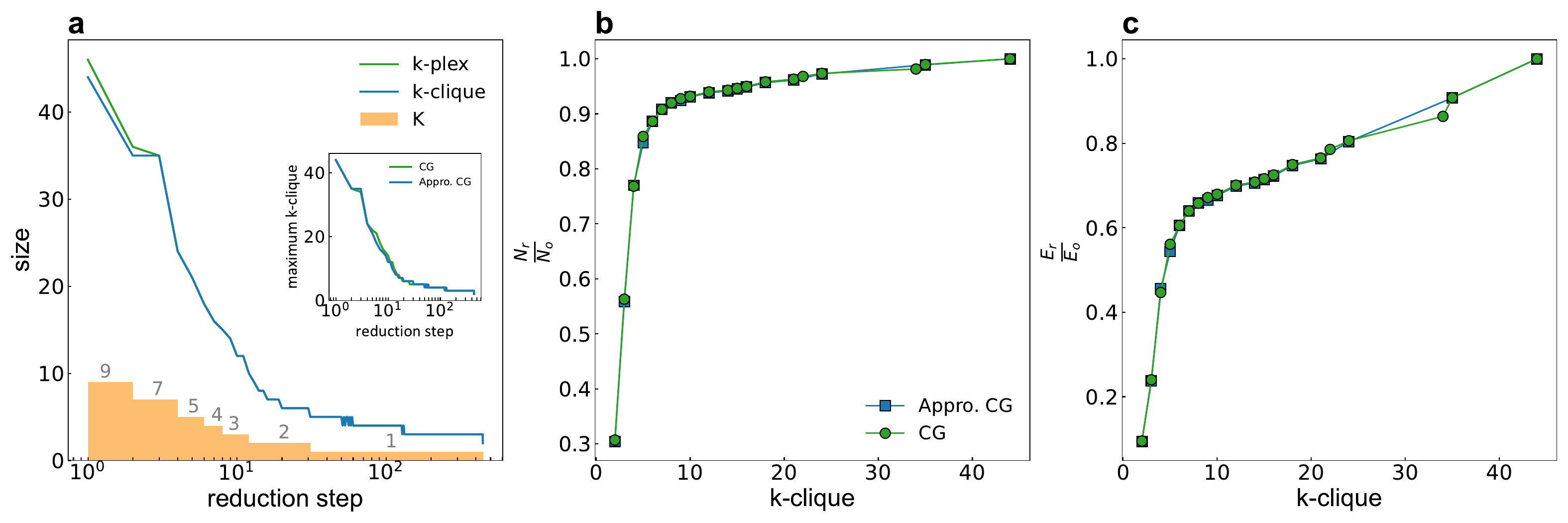}
	\caption{\textbf{Comparison of ISCG and approximate ISCG methods for network reduction}.
    \textbf{a} Maximum size of the clique and $k$-plex at each reduction step during the coarse-graining process.  
    The orange bars indicate the $k$-value for the $k$-plex structure at each reduction step, while the curve represent the results of ISCG~(clique-based) and approximate ISCG~(k-plex-based) methods.
    \textbf{b} Proportion of nodes~($N_r/N_o$) in the CGNs relative to original network.
    \textbf{c} Proportion of edges~($E_r/E_o$) in the CGNs relative to original network.
    These results, obtained using the GrQc network, demonstrate that approximate ISCG, which uses relaxed structural constraints (e.g., $k$-plexes instead of strict $k$-cliques), achieves comparable reduction performance to the original ISCG method.
    }
	\label{fig:figS4}
\end{figure*}

\label{SI:approximate}
In this section, we explore a relaxation of the structural constraints in the ISCG framework by utilizing $k$-plexes  in place of strict $k$-cliques during the coarse-graining process.
Unlike $k$-cliques, which require a fully connected structure, $k$-plexes are more flexible, allowing subgraph nodes to have up to $k$ missing edges. 
This relaxation reduces computational costs and enhances the adaptability of the framework for coarse-graining large-scale networks.

To balance computational efficiency and the preservation of contagion dynamics, 
the value of $k$ in the $k$-plex structure is adaptively determined based on the maximum clique size in the current network at each reduction step.
Specifically, we set $k=0.5\cdot m$, where $m$ denotes the maximum clique size at a given step.
This strategy ensures that the $k$-plex structure closely approximates the original clique structure while significantly simplifying the network.

Supplementary Figure~\ref{fig:figS4} compares the performance of the approximate ISCG method with the original ISCG framework on the GrQc network.
Both methods exhibit a similar reduction trajectory, with $k$-plex structures capturing larger subgraphs during the initial stages of reduction~(Supplementary Figure~\ref{fig:figS4} a). 
The adaptive nature of the method is highlighted by the orange bars,which indicate the $k$-values used for $k$-plexes. 
Early in the reduction, the $k$-plex approach identifies larger substructures, enabling faster simplification, while at later stages, the identified subgraphs increasingly resemble $k$-cliques (e.g., 1-plex).
This progression maintains contagion dynamics while optimizing structural reduction efficiency.

The proportions of nodes and edges retained in the coarse-grained networks relative to the original network are presented in Supplementary Figure~\ref{fig:figS4}b and c.
The approximate ISCG method achieves a reduction performance comparable to the original ISCG framework, yielding similar results across all reduction stages. 
Although the reduction effectiveness depends on the underlying network structure, the approximate approach demonstrates its ability to provide a practical alternative with minimal loss of accuracy or efficiency.

By leveraging the flexibility of $k$-plex structures, the approximate ISCG method effectively handles networks with diverse topological characteristics, making it particularly well-suited for coarse-graining large-scale networks.
The adaptability underscores its potential as a robust and efficient alternative for network reduction tasks, providing both structural simplification and dynamic preservation.

\subsection*{Contagion dynamic on CGNs} 
\label{SI:dynamics}
\begin{figure*}[!htbp]
	\centering
	\includegraphics[width=\textwidth]{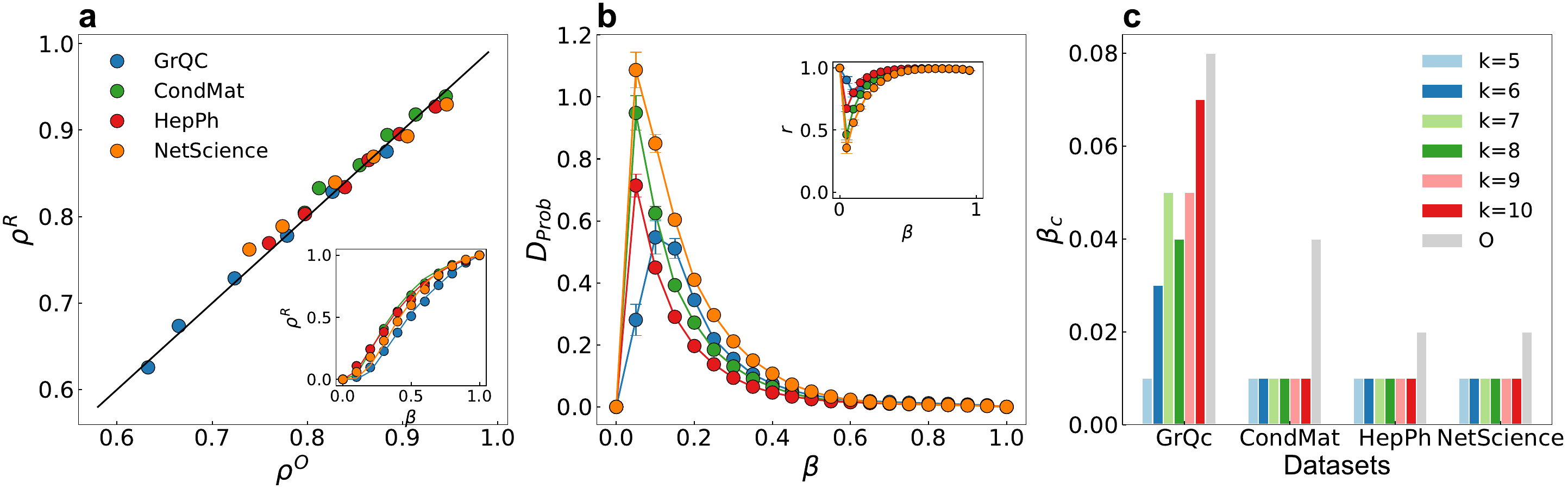}
	\caption{\textbf{Comparison of SIR dynamics between $k$-clique CGNs and the original network across four datasets: GrQc, CondMat, HepPh, NetScience}.
        \textbf{a} The final infected nodes density~($\rho^R$) on $k$-clique CGNs~(k = 5 to 10) versus original network~($\rho^O$) at $\beta=\hat{\beta}_k$. 
        The black solid line $y=x$ serves as a reference for equivalence.
        Inset: The final infected node density on $k$-clique CGNs plotted as a function of $\beta$.
        \textbf{b} Discrepancy~($D_{Prob}$) in infection probability between the original network and $5$-clique CGN as a function of $\beta$. 
        Inset: Pearson correlation coefficient~($r$) of infection probabilities between the 5-clique CGN and the original network. 
        \textbf{c} Critical infection thresholds~($\beta_c$) determined through numerical simulations using the susceptibility~($\chi$) for the original network and $k$-clique CGNs~($k$ = 5 to 10).
        }
	\label{fig:figS5}
\end{figure*}

In this section, we evaluate the performance of the ISCG framework in preserving the contagion dynamics of the original network using four datasets: GrQc, CondMat, HepPh, and NetScience. 
While the main manuscript  focuses on results from the GrQc network, this supplementary analysis provides a comprehensive assessment across all four datasets. 
To demonstrate the accuracy of the ISCG framework at various scales, we analyze the following aspects:
$\mathrm{(i)}$ Macroscopic scale: Accuracy in preserving the final infected node density ($\rho^R$). 
$\mathrm{(ii)}$ Microscopic scale: Discrepancy in the infection probabilities of nodes during contagion initiated from randomly selected seeds.
$\mathrm{(iii)}$ Critical phase transition: Precision in detecting the critical infection threshold ($\beta_c$).

Supplementary Figure~\ref{fig:figS5} a illustrates the agreement between the CGNs and the original network by comparing the final infected node density ($\rho^R$) for different $k$-values~(ranging from 5 to 10) at $\beta = \hat{\beta}_k$. 
The results show a strong correspondence~($\rho^R \approx \rho^O$), with points closely aligning to the reference line $y=x$. 
This consistency across datasets highlights the robustness of ISCG in accurately capturing macroscopic contagion dynamics.
The inset extends the analysis by exploring $\rho^R$ across a range of transmission probabilities $\beta$, beyond the theoretically derived minimum threshold $\hat{\beta}_k$.
The CGNs maintain high accuracy in capturing $\rho^R$ even at smaller $\beta$, with only minor deviations observed in the NetScience dataset.
This findings underscore the resilience of ISCG in preserving contagion dynamics under varying conditions.

At the microscopic level, Supplementary Figure~\ref{fig:figS5} b examines the discrepancies in node infection probabilities between the original network and the 5-clique CGN, using the metric $D_{Prob}$.
The discrepancy peaks at intermediate $\beta$, corresponding to the smallest infection probability that allows for significant contagion. 
As $\beta$ increases, $D_{Prob}$ decline rapidly, converging to zero around $\beta = 0.5$, where the CGNs accurately replicate the contagion dynamics of the original network.
The height of these peaks varies across datasets due to differences in network size.
For instance, NetScience, the largest dataset, exhibits the highest peak, while GrQc, the smallest, shows the lowest. 
Despite these variations, all curves converge as $\beta$ approaches $\hat{\beta}_k$, demonstrating that CGNs effectively preserve node-level contagion dynamics. 
The inset in Supplementary Figure~\ref{fig:figS5}b further illustrates the Pearson correlation coefficient ($r$) between infection probabilities in the CGNs and the original network.
The high correlation values~($r \rightarrow 1$) as $\beta$ increases confirm ISCG's capability to maintain microscopic contagion dynamics.

Supplementary Figure~\ref{fig:figS5} c assesses the ISCG's ability to determine the critical infection threshold ($\beta_c$) across multiple datasets.
While observed critical points are slightly lower than the theoretical $\hat{\beta}_k$, reflecting minor deviations inherent to the framework, they remain closely aligned with the actual thresholds of the original networks. 
These results demonstrate that $k$-clique CGNs reliably approximate the transition points in contagion dynamics.

Overall, these findings demonstrate the ISCG's effectiveness in preserving contagion dynamics at both macroscopic and microscopic scales, while achieving reasonable accuracy in identifying critical thresholds. 
The consistent performance across all four datasets underscores ISCG’s robustness and applicability for large-scale network analysis.

\begin{figure*}[!htbp]
	\centering
	\includegraphics[width=\textwidth]{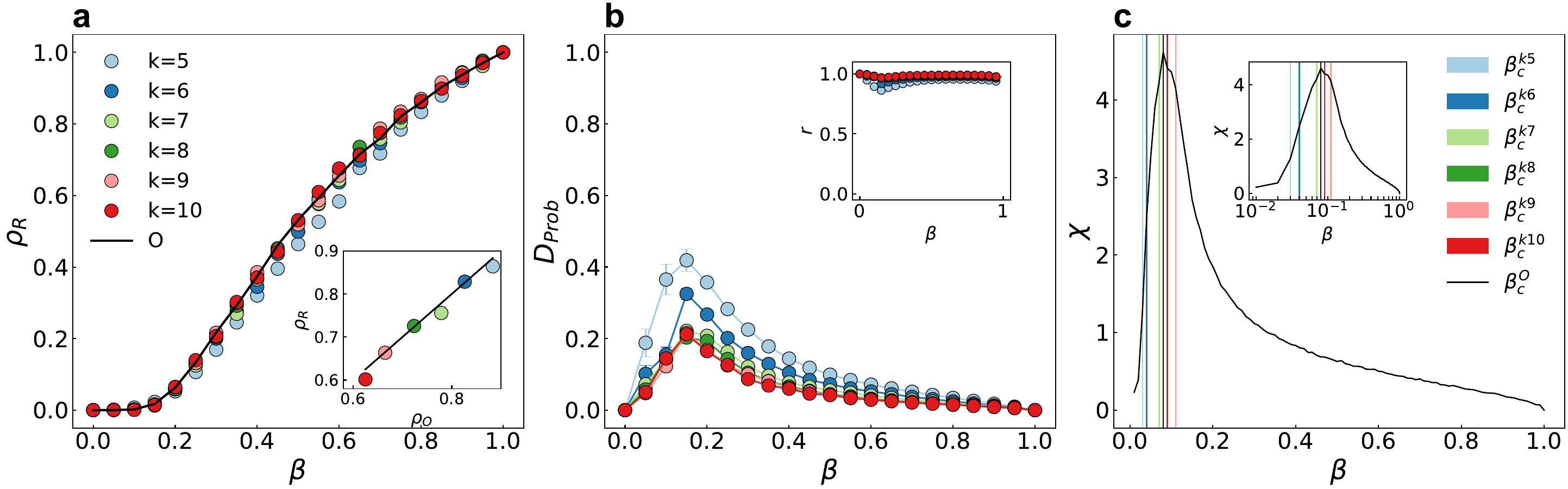}
	\caption{\textbf{Preservation of SIR dynamics on $k$-clique CGNs without edge weight information.}
        \textbf{a} The density of final infected nodes~($\rho_R$) as a function of $\beta$ for different $k$-clique CGNs~($k=5$ to 10).
        Inset: The comparison of $\rho_R$ on $k$-clique CGNs and the original network when $\beta=\hat{\beta}_k$.
        \textbf{b} The discrepancy~($D_{prob}$) in the probability of nodes being infected between the original network and $k$-clique CGNs as a function of $\beta$.
        Inset: The pearson correlation coefficient~($r$) of infection probabilities between the original network and various $k$-clique CGNs, demonstrating high correlation across different scales of $\beta$.
        \textbf{c} The critical infection threshold identified through numerical simulations for the original network and $k$-clique CGNs.
        Inset: A log-scale plot of $\beta_c$ is provided to highlight finer details and the consistency between the $k$-clique CGNs and the original network.
        All results presented in this figure are obtained from analyses performed on the GrQc network.
        }
	\label{fig:figS6}
\end{figure*}

\subsection*{Contagion dynamic on unweighted CGNs} 
\label{SI:unweight}

\begin{figure}[!htbp]
	\centering
	\includegraphics[width=0.45\textwidth]{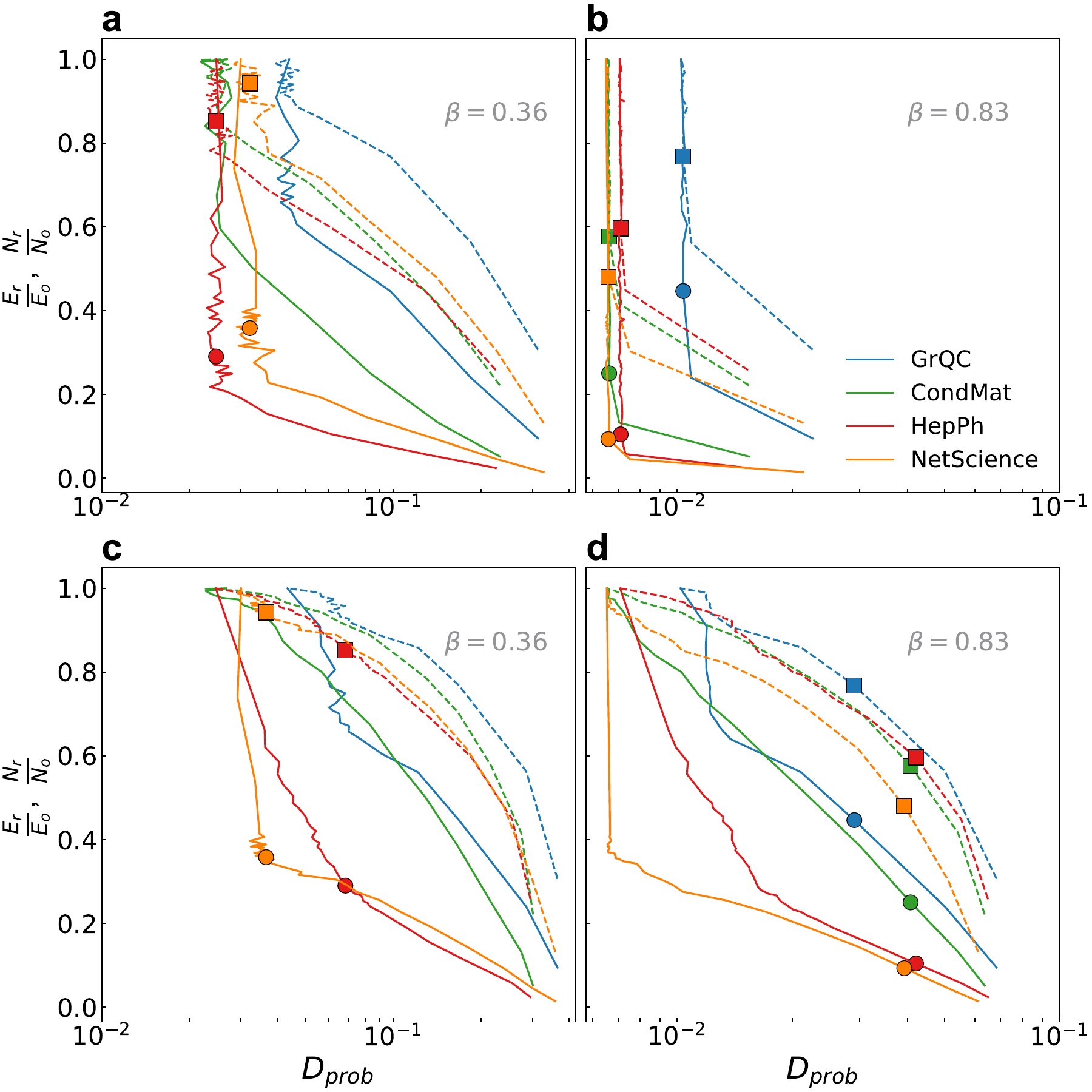}
	\caption{\textbf{Comparison of the balance between dynamic behavior accuracy and network reduction for weighted and unweighted CGNs}.
        The reduced ratios of nodes~($\frac{N_r}{N_o}$, dashed lines) and edges~($\frac{E_r}{E_o}$, solid lines) are shown as a function of the accuracy in preserving contagion dynamics~($D_{prob}$) for a given transmission probability.
        As the original network is progressively reduced to a single node, the plots depict the relationship between $D_{prob}$ and the corresponding reduction ratios of various $k$-clique CGNs.
        The transmission probability is set to $\beta=0.36$ in \textbf{(a, c)} and $\beta=0.83$ in \textbf{(b, d)}.
        These values correspond to the critical thresholds $\hat{\beta}_{10} = 0.36$ and $\hat{\beta}_3 = 0.83$, respectively.
        At these thresholds, the SIR contagion dynamics are exactly preserved when the original network is reduced to the 10-clique CGN (\textbf{a, c}) and 3-clique CGN~(\textbf{b, d}). 
        The positions of squares (node reduction) and circles (edge reduction) highlight the corresponding $k$-clique CGNs,  indicating high accuracy in maintaining the dynamic behavior.
        \textbf{(a, b)} Simulations consider weighted edges, capturing the effects of edge weights on the contagion process. 
        \textbf{(c, d)} Simulations are based solely on structural information, without edge weights, while still demonstrating the preservation of contagion dynamics.
        }
	\label{fig:figS7}
\end{figure}

Unweighted CGNs represent a simplified version of coarse-grained networks, retaining only the structural connections between super-nodes while disregarding the weights that encode connection strength. 
This approach offers several advantages: 
$\mathrm{(i)}$ Reduced computational complexity, enhancing the efficiency of analyzing large-scale networks.
$\mathrm{(ii)}$ Isolation of structural contributions, allowing for a focused evaluation of the network topology's role in contagion dynamics.
However, this simplification limits the granularity of the captured dynamics by neglecting connection strengths. 
By comparing unweighted and weighted CGNs, we assess the importance of edge weights in preserving contagion dynamics.
The results are presented in Supplementary Figures~\ref{fig:figS6} and \ref{fig:figS7}.

Supplementary Figure~\ref{fig:figS6}a compares the density of final infected nodes~($\rho^R$) for unweighted CGNs and the original network.
While unweighted CGNs exhibit slight deviations from the original network at $k = 5$, accuracy improves significantly as $k$ increases.
Despite the absence of edge weights, unweighted CGNs approximate key dynamic behaviors well, particularly at higher coarse-graining levels.

Supplementary Figure~\ref{fig:figS6}b examines discrepancies in infection probabilities~($D_{Prob}$)  between unweighted CGNs and the original network.
At intermediate $\beta$,  $D_{Prob}$ is relatively low but becomes more pronounced at higher $\beta$, reflecting the increased sensitivity of contagion propagation to edge weights.
As $\beta$ continues to rise, discrepancies diminish, demonstrating that structural topology dominates contagion dynamics at high infection probabilities.
The inset highlights a high Pearson correlation~($r \rightarrow 1$) between infection probabilities, reinforcing the capability of unweighted CGNs to preserve microscopic dynamics despite simplifications.

Supplementary Figure~\ref{fig:figS6}c evaluates the ability of unweighted CGNs to approximate the critical infection threshold~($\beta_c$).
The peaks in susceptibility $\chi$ align closely with the original network's critical points, with small deviations in $\beta_c$.
The results confirm that unweighted CGNs effectively capture critical phase transitions, particularly at higher $k$ values, where structural information compensates for the absence of edge weights.

Supplementary Figure~\ref{fig:figS7} provides a detailed comparison of weighted and unweighted coarse-grained networks (CGNs) in terms of their ability to maintain contagion dynamics accuracy and reduce network size.
This analysis examines two infection probabilities: a lower value ($\hat{\beta}_{10}=0.36$) and a higher value ($\hat{\beta}_{3}=0.83$).
As shown in Supplementary Figures~\ref{fig:figS7} a and b, the results indicate that for weighted CGNs, when the original network is reduced to the predicted scale at the given $\hat{\beta_k}$ (indicated by the markers), the curves exhibit a steep vertical drop in network complexity, while $D_{Prob}$ remains largely unaffected.
In contrast, for unweighted CGNs (Supplementary Figures~\ref{fig:figS7}c and d), the reduction curves show a more gradual decline at the predicted scale compared to the original network. 
This suggests that unweighted CGNs compromise some accuracy in preserving contagion dynamics to achieve network reduction.
For approximate reductions, where the network is further reduced beyond the predicted scale, weighted CGNs consistently outperform unweighted CGNs across all datasets and $\beta$ values.
This advantage is especially pronounced at higher $\beta$, where propagation dynamics are highly sensitive to connection strengths.
While unweighted CGNs achieve reasonable accuracy, incorporating edge weights captures the critical influence of connection strengths, resulting in superior performance in preserving contagion dynamics.
An intriguing observation arises when considering the trade-off between contagion accuracy and network complexity reduction. 
Weighted CGNs demonstrate a more substantial reduction in network complexity, as measured by the proportion of remaining nodes ($N_r / N_o$) and edges ($E_r / E_o$), 
while incurring minimal losses in contagion accuracy. Larger datasets, such as NetScience, exhibit even greater reductions, demonstrating the scalability of weighted CGNs for large-scale networks.
The results emphasize the robustness and efficiency of the weighted CGN framework, which achieves an optimal balance between accuracy and computational efficiency. 

\subsection*{Applications of the ISCG framework}
\label{SI:advantage}

\begin{figure*}[!htbp]
	\centering
	\includegraphics[width=\textwidth]{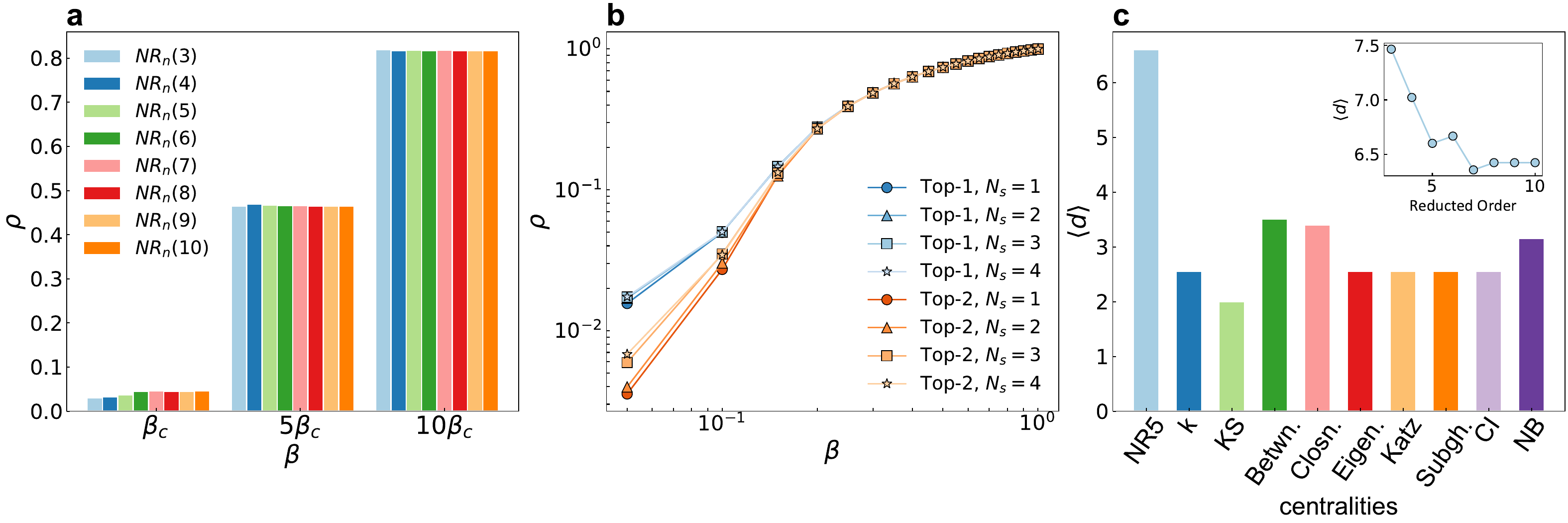}
	\caption{\textbf{Effectiveness of the ISCG-IM method in influence maximization}. 
    \textbf{a} The density of final infected nodes~($\rho$) with various $\beta$ values,  achieved by the ISCG-IM method across different $k$-clique CGNs ($k=3$ to $10$), using the top $n=10$ nodes as initial seeds for numerical simulations.
    This result highlights the robustness of ISCG-IM, exhibiting minimal sensitivity to the scale of the coarse-grained network.
    \textbf{b} The density of final infected nodes~($\rho$) in a contagion process initiated by multiple seeds~($N_s=1$ to $4$) randomly selected from within the same candidate set~(i.e., super-node). 
    Results are presented for the top two super-nodes with the largest weights in the 5-clique CGN.
    \textbf{c} The average shortest path $\langle d\rangle$ of the top 10 initial seeds selected by various methods. 
    Inset: The average shortest path $\langle d\rangle$ of the top 10 initial seeds identified by the ISCG-IM method on different $k$-clique CGNs~(k=3 to 10).
    All results are based on the GrQc network, revealing the factors contributing to the superior performance of the ISCG-IM method in maximizing spreading across multi-scale CGN networks.
    }
	\label{fig:figS8}
\end{figure*}

\textbf{ISCG-IM method}.
Supplementary Figure~\ref{fig:figS8}\textbf{a} examines the performance of the ISCG-IM method across a range of $k$-clique CGNs ($k = 3$ to $10$). 
The results reveal that the method is minimally affected by the choice of $k$, as the density of final infected nodes ($\rho$) remains consistent across different scales. 
This robustness ensures reliable performance regardless of the coarse-graining level, enabling the ISCG-IM method to adapt effectively to networks with diverse structures and resolutions.

Mechanisms behind effectiveness of ISCG-IM.
The superior performance of the ISCG-IM method in influence maximization, as highlighted in the main text, stems from its integration of macroscopic and microscopic insights.
By leveraging $k$-clique CGNs, the method identifies super-nodes that represent critical structural components, such as densely connected regions and their interconnections.
These coarse-grained representations provide macroscopic insights into the network's structure, enabling the evaluation of node influence from a global perspective.

This macroscopic view is complemented by localized seed selection within each super-node.
ISCG-IM avoids redundancy by distributing seeds across super-nodes, as shown in Supplementary Figure~\ref{fig:figS8}\textbf{b}.
When multiple seeds are placed within the same super-node, the final infected density ($\rho$) exhibits diminishing returns as $\beta$ increases.
Distributing seeds across super-nodes mitigates this overlap, maximizing the spread of influence. 
Within each super-node, the method refines seed selection using simple centrality metrics, such as degree, to identify the most influential nodes in densely connected regions.
This dual-scale approach---combining macroscopic structural insights with microscopic node-level features---ensures that ISCG-IM optimizes seed placement and maximizes the influence potential of selected seeds. 

Synergy among selected seeds.
To evaluate the collective influence of seeds identified by ISCG-IM, Supplementary Figure~\ref{fig:figS8}\textbf{c} analyzes the average shortest path between selected nodes.
The results show that ISCG-IM selects seeds with significantly longer average shortest paths compared to other methods.
This ensures that the seeds are well-dispersed across the network, minimizing redundancy and maximizing coverage to enhance overall effectiveness.
The inset in Supplementary Figure~\ref{fig:figS8}\textbf{c} 
explores the influence of $k$ on seed placement. 
As $k$ increases, the average shortest path between seeds decreases markedly, indicating more localized seed selection at higher coarse-graining levels. 
This observation highlights the importance of choosing an optimal $k$ to balance the benefits of coarse-grained representation with the need for well-distributed influence.

The ISCG-IM method's ability to integrate macroscopic community-level features with microscopic node-level influence underpins its robust and effective performance in influence maximization.
By synergizing these two scales, ISCG-IM outperforms adaptive centrality-based approaches, offering superior adaptability across multi-scale CGNs.
These findings underscore the versatility and potential of ISCG-IM for diverse applications in network propagation problems. 

\begin{figure*}[!htbp]
	\centering
	\includegraphics[width=\textwidth]{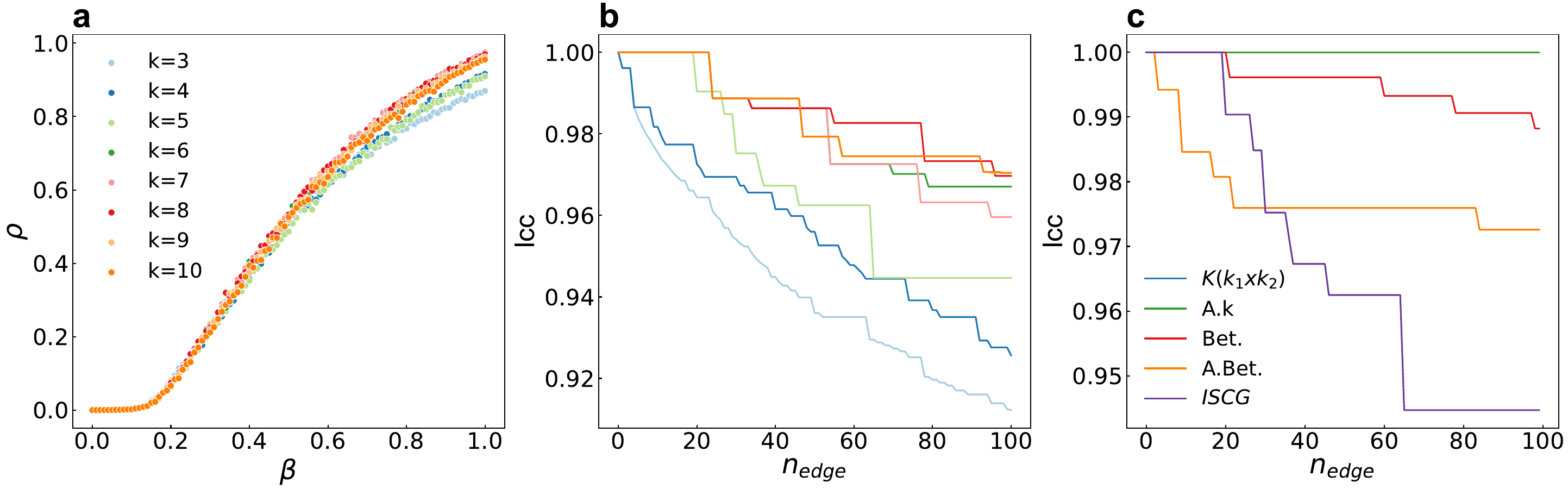}
	\caption{\textbf{Effectiveness of the ISCG-EI method in edge-based immunization}.
    \textbf{a} The density of final infected nodes $(\rho)$ achieved by the ISCG-EI method across various $k$-clique CGNs ($k=3$ to $10$).
    In the numerical simulations, 50 edges were immunized to evaluate the method's effectiveness.
    \textbf{b} The proportion of the largest connected component~(lcc) relative to the original network after the sequential removal of $n_{edge}$ edges,  identified by the ISCG-EI method across different $k$-clique CGNs.
    \textbf{c} Similar to \textbf{b}, but comparing the ISCG-EI method with other edge-based centrality metrics for edges removal.
    The comparison underscores the superior performance of the ISCG-EI~(k=5) method in diminishing the size of the largest connected component.
    All results are based on the GrQc network, showcasing the robustness and efficiency of the ISCG-EI method in  edge-based immunization. 
    }
	\label{fig:figS9}
\end{figure*}

\textbf{ISCG-EI method}.
Supplementary Figure~\ref{fig:figS9}\textbf{a} evaluates the robustness of ISCG-EI across various $k$-clique CGNs ($k = 3$ to $10$). 
The density of final infected nodes ($\rho$) remains consistent across varying $k$-values, demonstrating the method’s adaptability to different coarse-grained scales. Notably, at lower $k$ values (e.g., $k=3$), ISCG-EI shows slightly better performance, indicating that coarser representations enable the method to capture broader structural features that significantly influence infection pathways.

Structural impact on network connectivity.
To investigate the mechanisms behind ISCG-EI's effectiveness, Supplementary Figure~\ref{fig:figS9}\textbf{b} analyzes the largest connected component~(lcc) as a proportion of the original network size after removing $n_{edge}$ edges identified by ISCG-EI across $k$-clique CGNs. 
The results show that at coarser scales (e.g., $k=3$), the identified edges have a more pronounced impact on network connectivity, reflected by the steep reduction in the lcc proportion.
This highlights ISCG-EI’s ability to detect and disrupt structurally critical edges that hold the network together.
By targeting links pivotal to macroscopic connectivity, ISCG-EI effectively fragments the network, reducing its structural integrity.

Comparison with other edge-based methods.
Supplementary Figure~\ref{fig:figS9}\textbf{c} compares ISCG-EI with other edge-based centrality metrics, including adaptive betweenness and degree-based strategies, using the lcc metric. 
The results show demonstrate that ISCG-EI ($k=5$) consistently outperforms competing methods in reducing the largest connected component. 
This advantage arises from ISCG-EI's ability to prioritize edges whose removal triggers a cascading disruptions on network connectivity, fragmenting the network into smaller, disconnected components. 
In contrast, alternative methods often overlook such structurally critical edges in networks with highly localized dense connections, leading to less effective immunization.

ISCG-EI leverages structural insights from $k$-clique CGNs to identify edges essential for network connectivity and contagion dynamics. 
At coarser scales, the method captures high-level structural dependencies, allowing it to pinpoint edges with the greatest impact on both network integrity and infection spread. 
This adaptability makes ISCG-EI a powerful and versatile tool for immunization strategies across diverse networks and infection scenarios.

\begin{figure*}[!htbp]
	\centering
	\includegraphics[width=\textwidth]{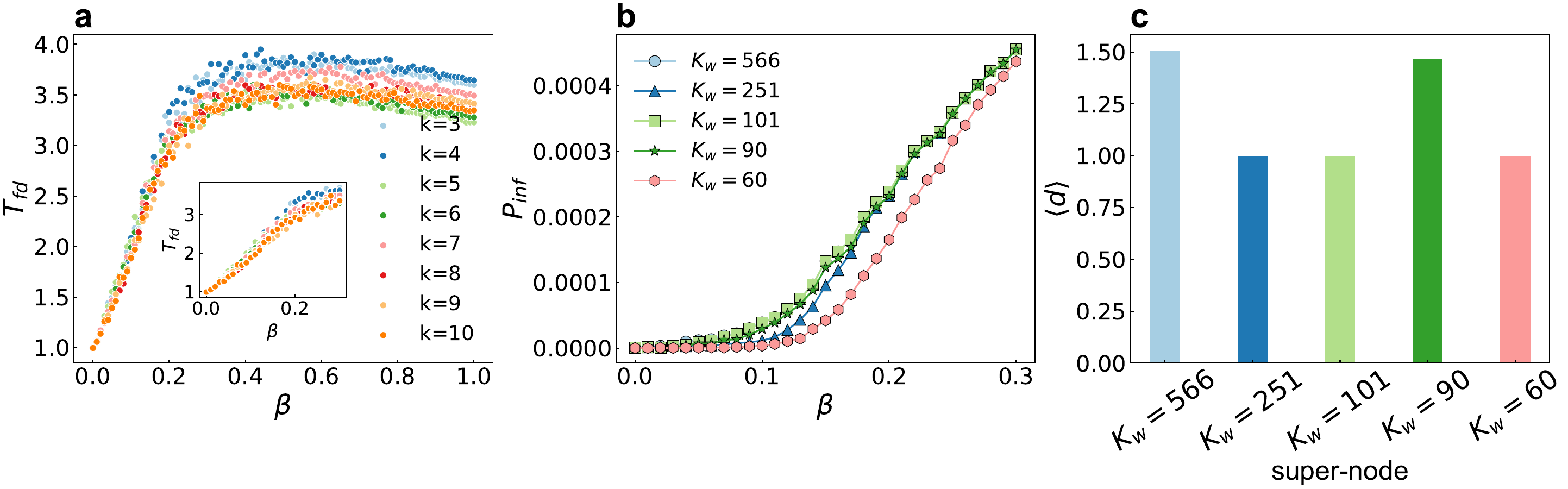}
	\caption{\textbf{Effectiveness of ISCG-SS in sentinel surveillance}.
    \textbf{a} The time to first detect the contagion signal~($T_{fd}$) using the top 10 sentinels nodes identified by the ISCG-SS across various $k$-clique CGNs~($k=3$ to $10$).
    Inset: A detailed view of $T_{fd}$ at smaller $\beta$ values, emphasizing the differences in detection time at low transmission probability.
    \textbf{b} The probability of super-nodes being infected ($P_{inf}$) in the 10-clique CGN is shown as a function of $\beta$. 
    The results focus on the top 5 super-nodes identified by ISCG-SS, where $K_w$ denotes the weighted degree of each super-nodes~(i.e. the sum of the weights of all edges connected to the super-node). 
    \textbf{c} The average shortest path length between nodes pair within individual super-nodes. 
    This metric offers insights into the structural compactness of super-nodes identified by ISCG-SS method. 
    All results are derived from the GrQc network, demonstrating the effectiveness of the ISCG-SS method  in enhancing sentinel surveillance by enabling early detection and ensuring robust monitoring capabilities.
    }
	\label{fig:figS10}
\end{figure*}

\textbf{ISCG-SS method}.
Supplementary Figure~\ref{fig:figS10}\textbf{a} evaluates the performance of ISCG-SS across $k$-clique CGNs ($k=3$ to $10$) using the GrQC network. 
The results show that $T_{fd}$ decreases with increasing $\beta$, with $k=5$ and $k=6$ achieving the smallest $T_{fd}$ values at higher $\beta$ levels. 
This demonstrates ISCG-SS's effectiveness in identifying sentinel nodes at intermediate coarse-grained scales, where structural and dynamic balances are best maintained. 
At smaller $\beta$ values, as shown in the inset, the differences across $k$ are minimal, indicating that the method's performance is less sensitive to the reduction scale in low-infection-rate regimes. 
This robustness across varying $k$ and $\beta$ underscores ISCG-SS’s adaptability to diverse network configurations and epidemic scenarios.

Mechanisms behind ISCG-SS effectiveness.
To explore the mechanisms driving ISCG-SS’s superior performance, Supplementary Figures~\ref{fig:figS10}\textbf{b} and \textbf{c} analyze two critical structural properties of the selected sentinel nodes.
Super-node infection probability: Supplementary Figure~\ref{fig:figS10}\textbf{b} examines the probability of super-nodes being infected ($P_{inf}$) as a function of $\beta$ for the top 5 super-nodes identified in the 10-clique CGN. 
The results reveal a positive correlation between the weighted degree ($K_w$) of a super-node and its likelihood of infection. 
Super-nodes with higher $K_w$ values are more exposed to incoming infections due to their extensive connections, making sentinel nodes from these super-nodes highly effective at detecting outbreaks early by intercepting contagion pathways.

Internal connectivity of super-nodes: Supplementary Figure~\ref{fig:figS10}\textbf{c} evaluates the average shortest path length $\langle d \rangle$ within individual super-nodes, reflecting their structural compactness.
This results show minimal variation in $\langle d \rangle$ across super-nodes with different $K_w$ values, indicating  tightly connected internal structures in all cases. 
This ensures that once a sentinel node within a super-node is infected, the infection rapidly spreads to other nodes within the same super-node. 
Combined with the insights from Supplementary Figure~\ref{fig:figS10}\textbf{b}, these findings highlight how ISCG-SS integrates macroscopic (super-node level) and microscopic (intra-super-node) features to achieve efficient outbreak detection.

The ISCG-SS method demonstrates robust and efficient performance in sentinel surveillance by integrating coarse-grained structural insights with dynamic properties of infection spread. 
At higher transmission probability, intermediate coarse-grained scales ($k=5$ and $k=6$) strike a balance between structural simplification and dynamic fidelity, enabling optimal sentinel selection. 
Furthermore, the ability to target super-nodes with high infection probabilities and tightly connected internal structures ensures rapid and reliable outbreak detection.
These findings establish ISCG-SS as a powerful and adaptable tool for enhancing sentinel surveillance in diverse networks and propagation scenarios.

\end{document}